\documentclass[fleqn,usenatbib]{mnras}

\usepackage{newtxtext,newtxmath}

\usepackage[T1]{fontenc}

\DeclareRobustCommand{\VAN}[3]{#2}
\let\VANthebibliography\thebibliography
\def\thebibliography{\DeclareRobustCommand{\VAN}[3]{##3}\VANthebibliography}

\usepackage{graphicx}	
\usepackage{amsmath}	
\usepackage{gensymb}
\usepackage[table]{xcolor}    
\usepackage{pdflscape}
\usepackage{afterpage}
\usepackage{capt-of}
\usepackage{longtable}
\usepackage{multirow} 
\usepackage{subcaption}
\usepackage{breqn}

\title[A Study of Flares in the Ultra-Cool Regime from SPECULOOS-South]{A Study of Flares in the Ultra-Cool Regime from SPECULOOS-South}

\author[C. A. Murray et al.]{
C. A. Murray$^{1}$\thanks{E-mail: cam217@cam.ac.uk}, 
D. Queloz$^{1}$,
M. Gillon$^{2}$,
B. O. Demory$^{3}$,
A. H. M. J. Triaud$^{4}$,
J. de Wit$^{5}$, \newauthor
A. Burdanov$^{5}$,
P. Chinchilla$^{2,6}$,
L. Delrez$^{2,7}$,
G. Dransfield$^{4}$,
E. Ducrot$^{2}$, 
L. J. Garcia$^{2}$, 
\newauthor
Y. G\'omez Maqueo Chew$^{8}$,
M. N. G{\"u}nther$^{9}$\thanks{Juan Carlos Torres Fellow}, 
E. Jehin$^{7}$,
J. McCormac$^{10}$,
P. Niraula$^{5}$,
\newauthor
P. P. Pedersen$^{1}$, 
F. J. Pozuelos$^{2,10}$,
B. V. Rackham$^{5,9}$\thanks{51 Pegasi b Fellow}, 
N. Schanche$^{3}$,
D. Sebastian$^{4}$, 
\newauthor
S. J. Thompson$^{1}$, 
M. Timmermans$^{2}$,
R. Wells$^{3}$
\\
$^{1}$Cavendish Laboratory, JJ Thomson Avenue, Cambridge CB3 0HE, UK\\
$^{2}$Astrobiology Research Unit, University of Li\`ege, All\'ee du 6 ao\^ut, 19, 4000 Li\`ege (Sart-Timan), Belgium \\
$^{3}$University of Bern, Center for Space and Habitability, Gesellschaftsstrasse 6, 3012 Bern, Switzerland \\
$^{4}$School of Physics and Astronomy, University of Birmingham, Edgbaston, Birmingham B15 2TT, United Kingdom \\
$^{5}$Department of Earth, Atmospheric and Planetary Sciences, MIT, 77 Massachusetts Avenue, Cambridge, MA 02139, USA \\
$^{6}$ Instituto de Astrof\'isica de Canarias (IAC), Calle V\'ia L\'actea s/n, 38200, La Laguna, Tenerife, Spain\\
$^{7}$Space sciences, Technologies and Astrophysics Research (STAR) Institute, University of Li\`ege, Belgium\\
$^{8}$ Instituto de Astronom\'ia, Universidad Nacional Aut\'onoma de M\'exico, Ciudad Universitaria, Ciudad de M\'exico, 04510, M\'exico \\
$^{9}$Department of Physics, and Kavli Institute for Astrophysics and Space Research, Massachusetts Institute of Technology, Cambridge, MA 02139, USA\\
$^{10}$Department of Physics, University of Warwick, Coventry, CV4 7AL, UK
}

\date{Accepted XXX. Received YYY; in original form ZZZ}

\pubyear{2021}

\begin{document}
\label{firstpage}
\pagerange{\pageref{firstpage}--\pageref{lastpage}}
\maketitle

\begin{abstract}
We present a study of photometric flares on 154 low-mass ($\leq0.2\textrm{M}_{\sun}$) objects observed by the SPECULOOS-South Observatory from 1st June 2018 to 23rd March 2020. In this sample we identify 85 flaring objects, ranging in spectral type from M4 to L0. We detect 234 flares in this sample, with energies between $10^{29.2}$ and $10^{32.7}$\,erg, using both automated and manual methods. With this work, we present the largest photometric sample of flares on late-M and ultra-cool dwarfs to date. By extending previous M dwarf flare studies into the ultra-cool regime, we find M5--M7 stars are more likely to flare than both earlier, and later, M dwarfs. By performing artificial flare injection-recovery tests we demonstrate that we can detect a significant proportion of flares down to an amplitude of 1 per cent, and we are most sensitive to flares on the coolest stars. Our results reveal an absence of high-energy flares on the reddest dwarfs. To probe the relations between rotation and activity for fully convective stars, we extract rotation periods for fast rotators and lower-bound period estimates of slow rotators. These rotation periods span from 2.2\,hours to 65\,days, and we find that the proportion of flaring stars increases for the very fastest rotators. Finally, we discuss the impact of our flare sample on planets orbiting ultra-cool stars. As stars become cooler, they flare less frequently; therefore, it is unlikely that planets around the very reddest dwarfs would enter the `abiogenesis' zone or drive visible-light photosynthesis through flares alone.
\end{abstract}




\section{Introduction}
In the search for planets capable of supporting life, ultra-cool dwarfs (UCDs) make compelling hosts. In our local stellar neighbourhood UCDs are plentiful \citep{Chabrier2003GalacticFunction, Henry2004TheEnd} and predicted to host large numbers of planets \citep{Cantrell2013TheNeighbors, Dressing2015THESENSITIVITY,Ballard2016PLANETS}. Moreover, their small sizes and low temperatures make it easier to detect Earth-sized, habitable-zone planets around these objects, and to probe those planets' atmospheres for biosignatures \citep{Kaltenegger2009TransitsPlanets, Seager2009TransitingJWST, deWit2013ConstrainingSpectroscopy}, than for any other type of star. 

Despite their promise, serious questions remain about the habitability of planets around red dwarfs. Several authors find it unlikely that these extremely cool, red stars would provide their planets with enough UV photons to initiate specific prebiotic chemistry pathways \citep{Rimmer2018TheExoplanets} or enough visible light for photosynthesis to occur \citep{Mullan2018PhotosynthesisFlares,Lehmer2018TheStars, Covone2021EfficiencyZone}. UCDs are also especially active objects \citep{West2015AnDwarfs,Williams2015SimultaneousJ13142039+1320011, Gizis2017Superflare, Paudel2018Dwarfs, Gunther2020Dwarfs}, producing energetic stellar flares and large-scale photometric variability, which results in treacherously variable conditions for their planets. The dynamic relationship between a planet and its host is a major factor in evaluating how conducive a planet's environment is for life. For UCDs, however, this relationship is poorly understood.

Stellar flares may be a major determinant in whether a planet can initiate and sustain life. Flares are explosive events caused by magnetic recombination in the upper atmosphere of a star \citep{Benz2010PhysicalObjects}. This sudden eruption of magnetic energy usually occurs around active regions, such as stellar spots, and causes bursts of particles and electromagnetic radiation. The spectra of this radiation resembles a black body with an effective temperature of 9000\,K \citep{Shibayama2013SuperflaresSuperflares}. Often, though not always, powerful flares will come accompanied with a coronal mass ejection (CME) event, where clouds of charged particles are directionally ejected from the star. While flares can be destructive, through atmospheric erosion \citep{Lammer2007CoronalZones}, ozone depletion \citep{Segura2010TheDwarf, Tilley2019ModelingPlanet}, and even extinction events, they can also be an essential power source for life. It is possible that flares could provide the missing energy at the bluer end of the spectrum needed for cool, red dwarfs to initiate prebiotic chemistry \citep{Buccino2007UVStars, Ranjan2017TheFollow-up,Rimmer2018TheExoplanets} and for photosynthesis \citep{Mullan2018PhotosynthesisFlares,Lingam2019PhotosynthesisStars}. The additional UV energy may also affect the evolution of a planet's atmospheric chemistry \citep{Segura2010TheDwarf, Vida2017FrequentLife}. In summary, stellar flares have extreme and far-reaching consequences on the planets hosted by cool stars. Therefore, it is essential to study these stars' flaring activity, and assess how applicable our current understanding of stellar activity is to the ultra-cool regime.

Stellar activity and rotation are also closely connected. Magnetised stellar winds dictate the loss of angular momentum, which slows a star's rotation over time. These stellar winds are dependent on the structure and properties of the magnetic field. As the rotation slows, this decreases the magnetic activity in a process known as spin down \citep{Skumanich1972TIMEDEPLETION, Noyes1984RotationStars.}. Due to this effect, rapidly rotating stars flare much more frequently than slow rotators \citep{Skumanich1986SOMESTARS, Davenport2019TheAge, Mondrik2018AnDwarfs, Medina2020FlarePc}. Spots and faculae on the surface of stars and brown dwarfs cause periodic photometric variations as they come in and out of view, on the same timescale as the object's rotation. This allows the rotation period to be deduced directly from the photometry. Therefore, the rotation of a star can provide valuable insights into the magnetic dynamo, the mechanism which generates a star's magnetic field.

This magnetic dynamo is poorly constrained for fully convective low-mass objects (with masses $\leq0.35\textrm{M}_{\sun}$, \citet{Chabrier1997StructureStars}), where it is believed to differ significantly from the solar model. Despite this predicted difference, recent work has shown that relationships between activity and rotation remain consistent from partially to fully convective stars \citep{Wright2016Solar-typeTachocline, Newton2017ROTATION, Wright2018TheDwarfs}. Spin down is, however, believed to occur on slower timescales for fully convective stars, which accounts for the enhanced activity of mid-to-late M dwarfs \citep{West2008CONSTRAININGSAMPLE, Newton2017ROTATION, Jackman2021StellarSurvey}. Therefore, rotation, activity, and the relationship between the two, are extremely useful probes of the underlying magnetic activity of ultra-cool dwarfs.

Due to the promising nature of M dwarfs as planetary hosts, there have been several detailed studies of their flaring activity within the past decade. Space telescopes, such as the Kepler/K2 missions \citep{Borucki2010KeplerResults,Howell2014TheResults}, allowed the first insights into the flares of bright M dwarfs \citep{Davenport2014Kepler1243, Hawley2014KeplerDwarfs, Lurie2015KeplerB, Silverberg20161243} and of (small numbers of) ultra-cool dwarfs \citep{Paudel2018Dwarfs, Gizis2013KeplerFlares, Gizis2017Superflare}. Additionally, the recently launched Transiting Exoplanet Survey Satellite (TESS, \citet{Ricker2015TransitingTESS}) has facilitated studies of the flaring and rotating activity of cool stars \citep{Gunther2020Dwarfs,Medina2020FlarePc,Seli2021ActivityTESS}. Several ground-based photometric surveys, such as MEarth \citep{Nutzman2008DesignDwarfs}, the Next Generation Transit Survey (NGTS, \citet{Wheatley2018TheNGTS}), the All-Sky Automated Survey for Supernovae (ASAS-SN, \citet{Shappee2014THE2617}) and Evryscope \citep{Law2015EvryscopeTelescopes} have also carried out detailed flare studies that include M dwarfs \citep{West2015AnDwarfs,Mondrik2018AnDwarfs,Jackman2021StellarSurvey,Schmidt2019TheASAS-SN, Martinez2019AASAS-SN,Howard2019EvryFlare.Sky, Howard2020EvryFlare.Sky, Howard2020EvryFlare.TESS}. 

The most frequent stellar flares are small, fast and difficult to detect above photometric scatter \citep{Lacy1976UVData.}. Due to the intrinsic faintness of UCDs in the visible, it is difficult to achieve the high photometric precisions necessary to constrain their flaring activity. However, it is possible to perform small, dedicated studies of the much rarer, high-energy flares on UCDs \citep{Gizis2013KeplerFlares,Paudel2018Dwarfs,Jackman2019DetectionSurvey}. Therefore, to obtain a sufficient sample size, previous large flare studies have focused on hotter stars, up to mid-M dwarfs, where photometric precisions are much higher. This has resulted in limited flare statistics for the coolest stars, for which we would require a large, high-cadence photometric survey optimised for UCDs.

The SPECULOOS (Search for habitable Planets EClipsing ULtra-cOOl Stars) project \citep{Gillon2018SearchingWorlds,Burdanov2018SPECULOOSTRAPPIST,Delrez2018SPECULOOS:Dwarfs,Jehin2018ThePlanets,Sebastian2021SPECULOOS:Strategy} aims to search for transiting planets around the nearest (within 40\,pc) ultra-cool dwarf stars. SPECULOOS's motivation is to provide temperate, terrestrial planets for detailed atmospheric characterisation with the James Webb Space Telescope (JWST, \cite{Gardner2006TheTelescope}) and future extremely large telescopes. The SPECULOOS target catalogue of ultra-cool objects is defined in \cite{Sebastian2021SPECULOOS:Strategy}. SPECULOOS comprises a network of 1-m class telescopes spread across the Northern and Southern Hemispheres. The largest facility, the SPECULOOS-Southern Observatory (SSO) in Cerro Paranal, Chile, consists of four identical, robotic telescopes. These telescopes are named Io, Europa, Callisto and Ganymede. Each telescope operates independently and in robotic mode following plans written by SPECULOOS’s automatic scheduler, SPeculoos Observatory sChedule maKer (\textsc{spock}, \citealt{Sebastian2021SPECULOOS:Strategy}). Additionally, the SPECULOOS-Northern Observatory \citep{Niraula2020Team} in Tenerife and the SAINT-EX (Search And
characterIsatioN of Transiting EXoplanets) facility \citep{Demory2020ATOI-1266} in San Pedro M\'artir each have one 1-m telescope. The SSO began official scientific operations in January 2019, followed by SAINT-EX in March 2019 and the SNO in June 2019.

In this paper we present a study of flares and rotation of SSO targets, observed over a span of almost two years. Section \ref{ssosample} defines the SSO data sample. Section \ref{methods} outlines how we generate and clean our global lightcurves and how we combine an automated flare detection algorithm with manual vetting to obtain our final flare sample. We describe modelling the flares, calculating the flare energies, and estimating flare rates. This section also includes an assessment of the flare sample's completeness. In Section \ref{rotation} we measure the rotation periods in our sample. The results of our flare and rotation analyses are presented in Section \ref{results} and discussed in Section \ref{discussion}, which includes a contextualisation of the impact of flares on the potential for life on planets around ultra-cool dwarfs.




\section{SPECULOOS-South Data Sample}\label{ssosample}
For this study we defined a dataset spanning from 1st June 2018 to 23rd March 2020, using observations from all four telescopes in the SSO. While this start date is before official scientific operations began, it marks a point of stability in the commissioning phase. After this date we performed no major maintenance to the observatory; DONUTS \citep{McCormac2013DONUTS:Stars}, our auto-guiding software for precise pointing, was in use; and our operating strategy had been finalised (see \citet{Sebastian2021SPECULOOS:Strategy} for details). The end of the data package was defined as the date that ESO Paranal Observatory shutdown due to the coronavirus pandemic.

In the time period between 1st June 2018 and 23rd March 2020, there are 661 potential nights of observation. Ganymede, the final telescope installed in the SSO, started commissioning on 30th September 2018, therefore, between the four telescopes there is a cumulative total of 2523 nights. However, primarily due to weather loss, the SSO has lost 23--24 per cent of observing time, resulting in 1931 nights of observation. Over the 1931 (combined) nights in this sample, we have observed 176 unique photometric targets for at least one night. These observations have typical exposure times of 20--60 seconds. The majority, though not all, of this data sample are in the SPECULOOS target list (86 per cent), as defined in \cite{Sebastian2021SPECULOOS:Strategy}.

This target list is divided into three main scientific programs, which are described in more detail in \citet{Sebastian2021SPECULOOS:Strategy}. The objects in the SSO sample which are not in the target list are exclusively objects with spectral types around or earlier than M6. It is likely these objects were part of the commissioning phase of the telescopes or were removed from the target list (due to reclassifying their spectral type) before official scientific operations began. For the objects in the target list, we extract radii, masses, effective temperatures, and spectral type classifications from \cite{Sebastian2021SPECULOOS:Strategy}. The stellar parameters for non-target-list objects are calculated in the same way.

We remove any objects that are not part of SPECULOOS's usual ``survey mode'', such as targets observed for follow-up and monitoring of TRAPPIST-1's transits. We note that in SPECULOOS's ``survey mode'', there are no simultaneous observations with multiple telescopes during the course of a night. During this time period the operational temperature of the CCD was increased from $-$70 to $-$60$\degree$C in October 2018. The choice to raise the temperature of the CCD was due to the effect on the quantum efficiency of the detector, improving our sensitivity at the red limit, while the increase in dark current was found to be negligible. This temperature change introduces an offset in the differential flux between the nights before and after. Whilst this offset has no consequence on our flare program, it affects long-term photometric trends. Therefore, when recovering rotation periods (in Section \ref{rotation}), we split the lightcurves that straddle this temperature change and analyse the before and after sections independently. As this temperature change happens during the first few months of our dataset, the majority of targets were either observed entirely before or after it. Therefore, we only need to split a handful of our lightcurves. If a target has been observed by multiple telescopes we do not combine their lightcurves, as each will experience slightly different instrumental systematics. However, we do check that the rotation periods estimated in Section \ref{rotation} are present in observations from all telescopes.

\begin{figure}
	\includegraphics[width=\columnwidth]{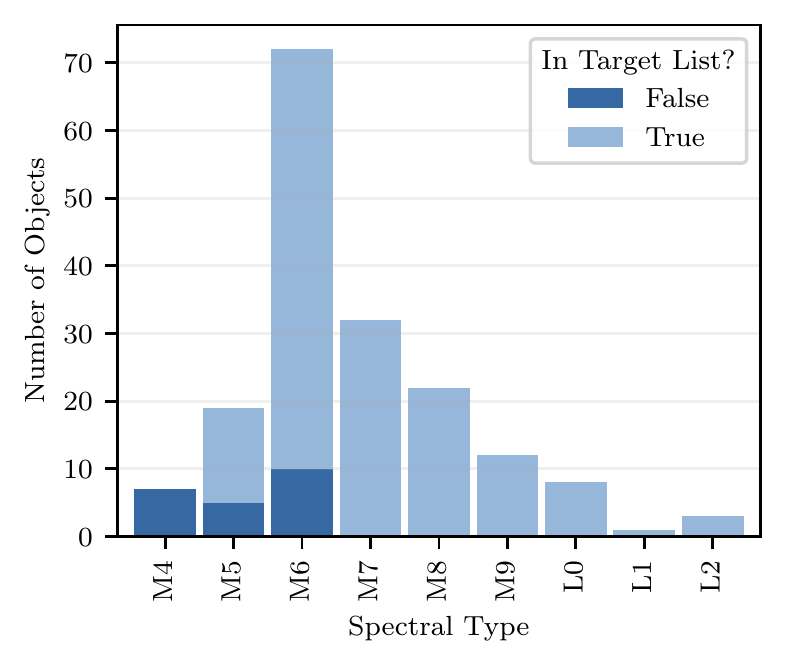}
    \caption{Spectral type distribution of the objects in the SPECULOOS-South data sample from 1st June 2018 to 23rd March 2020. We indicate which objects are in the SPECULOOS target list \citep{Sebastian2021SPECULOOS:Strategy}.}
    \label{fig:sso_sample}
\end{figure}

If we choose only the objects which have been observed more than 20 hours with one telescope, this provides us instead with 154 targets (of which 134, or $\sim$87 per cent, are in the SPECULOOS target list). We define the observation time as the sum of the span of all observation nights (start of night to end of night), where we exclude any gap longer than 15 minutes. These 154 objects define the SPECULOOS-South data sample. This sample covers a range of M dwarfs in spectral type, extending from M4 into the early L dwarf regime (with masses of 0.07--0.2\,$\textrm{M}_{\sun}$), as shown in Figure \ref{fig:sso_sample}. All objects in this sample will therefore be fully convective, as the convection limit occurs around 0.35\,$\textrm{M}_{\sun}$ \citep{Chabrier1997StructureStars}. While M4 and M5 objects are not considered ultra-cool dwarfs, we include them in this sample to explore any differences between mid-M, late-M, and L dwarfs. 

\begin{figure}
	\includegraphics[width=\columnwidth]{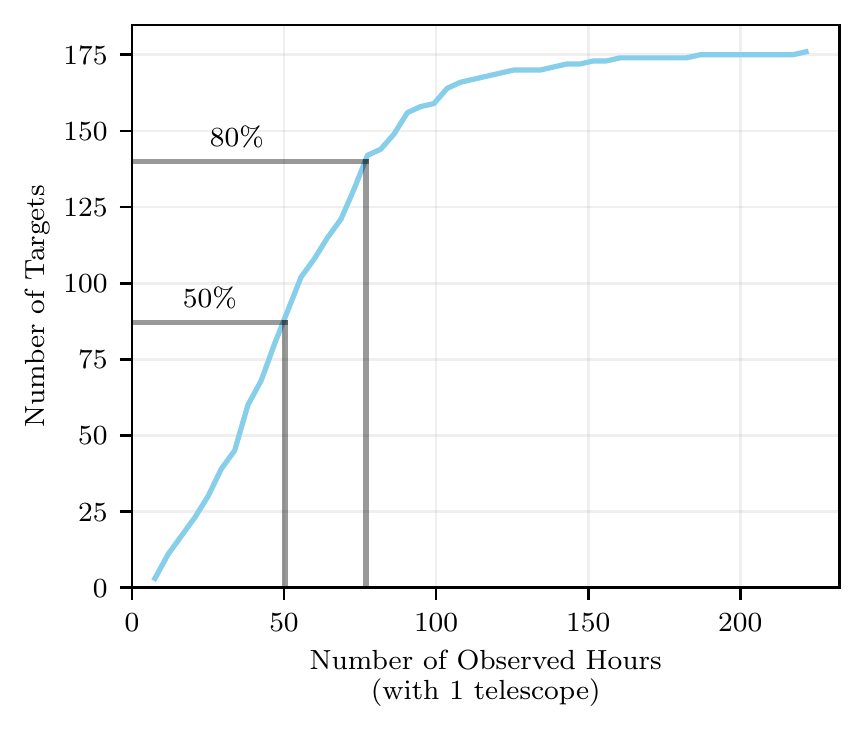}
    \caption{A cumulative histogram of the number of days observed for the objects in the SSO data sample. Of the 176 objects, 14 per cent (24) have been observed for less than 20 hours, 50 per cent have been observed for less than 51 hours, and 80 per cent have been observed for less than 77 hours.}
    \label{fig:sso_sample_hist}
\end{figure}

In this sample, 80 per cent of the objects have been observed for less than 77 hours (Figure \ref{fig:sso_sample_hist}), and 50 per cent have been observed for less than 51 hours. The short observation times in this sample will inevitably bias us towards detecting flares on targets with high flare rates in Section \ref{flaredetection} and detecting the clearest rotation periods for fast rotators in Section \ref{rotation}.

From \citet{Sebastian2021SPECULOOS:Strategy} we confirm that none of the target list objects in the SSO data sample are known binaries. We could not find any record in the literature of binarity for any of the 20 non-target list objects, except from the M4 star, Gaia DR2 4450376396936878336 \citep{Lepine2007Systems}. We flag this object, however, we find no evidence that this object behaves anomalously. This star is flaring, however, it only flares twice, therefore, it is not included in our analysis in Section \ref{ffd}. We also do not find a rotation period for this object, therefore it does not impact our results in Section \ref{rotation}.



\section {Generating the SPECULOOS-South Flare Sample}\label{methods}

\subsection{Global lightcurves}\label{globallcs}

To search for flares, we extract global lightcurves for all targets in the data sample, using the SSO Pipeline described in \cite{Murray2020PhotometrySPECULOOS-South}. The SSO Pipeline performs automated differential photometry by generating an `artificial lightcurve' from a normalised, weighted average of reference stars in the field. Only objects above a brightness threshold (optimised for each field) are chosen as reference stars. Reference stars whose differential lightcurves exhibit photometric variability, or that are far in spatial distance on-sky from the target, are then weighted down in an iterative algorithm, based on \citet{Broeg2005AStar}. It is worth noting that the target star is removed from this process. This means that, except for the distance weighting, this `artificial lightcurve' is entirely independent of the target star's lightcurve. This pipeline then corrects for the second-order effects of atmospheric water absorption on our differential photometry, using measurements of precipitable water vapour from the ground-based LHATPRO (Low Humidity and Temperature PROfiling radiometer) instrument \citep{Kerber2012AObservatory}.

To generate global lightcurves, we perform our differential photometry process on the entire time series of a target at once. We obtain differential lightcurves for 13 different aperture sizes. For each aperture, the aperture size, comparison stars, and weightings do not change over the time span of observations. Using global lightcurves allows us to study their long-term photometric variability, but restricts our ability to optimise the photometry night-by-night.

To mitigate any systematics caused by changing atmospheric conditions, we carefully select the `best' aperture for each target. As we only use one aperture size across all nights, it is important to take time at this step. Choosing the best aperture for a target involves a fine balance between an aperture that's large enough to avoid losing flux for nights with sub-optimal observing conditions (e.g., a larger seeing), and an aperture that's small enough to prevent blending from nearby stars and extra white-noise contamination from the background. We select the aperture by eye which minimises the correlation between the lightcurve flux and the seeing for the whole lightcurve, and has no clear contamination from neighbouring stars. In addition, we implement a bad weather flag and thresholds for the background sky level, airmass, and seeing (above which the data are severely impacted), described in more detail below.

\subsection{Cleaning the lightcurves}
\subsubsection{Removing `bad' observations}
In the context of this paper, `bad' observations are defined as those which are significantly affected by the observing conditions of the night, where distinguishing real stellar variability from ground-based systematics would be extremely difficult.

To reduce the impact of these observations on our flare study, we implement a bad weather flag, as defined in \cite{Murray2020PhotometrySPECULOOS-South}. As previously mentioned, the artificial lightcurve has no relation to the target's lightcurve, therefore it is not impacted by flares on the target star. This distinction means that we can use the artificial lightcurve as an independent reflection of the photometric conditions at that time to identify where the quality of our lightcurves deteriorates. Data points in our lightcurves are flagged based on the RMS of the section of the artificial lightcurve surrounding that data point (local RMS). We define the length of the local section to be $\pm0.01$d. We make the assumption here that there is a threshold for photometric scatter in the artificial lightcurve (which therefore must be present in the highly weighted reference stars' lightcurves) above which we cannot obtain good photometric precision. We define this threshold as a local RMS of 8 per cent, as defined in \citet{Murray2020PhotometrySPECULOOS-South}.

We also include strict cuts on specific observation parameters. We exclude data points in our lightcurves where the sky background level is greater than 4000 counts per pixel, the airmass is greater than 2.5, or the full width at half maximum (FWHM) of the point spread function (PSF) is greater than 2.6\,arcsec (the SSO cameras have a pixel scale of 0.35\,arcsec\,pixel$^{-1}$).

From 1st May to 19th June 2019 we experienced an issue with moving dust on the CCD window of Ganymede. While stationary dust can be easily corrected by flat images taken at the start of each night, if dust moves across the CCD window during the night, this leads to residuals in the flat correction and structures in the final differential lightcurves. These structures can mimic planetary transits, though the frames affected by dust are easy to identify from the raw images. However, as we do not currently have a robust correction technique for moving dust, we removed all observations taken by Ganymede during this period. This amounts to less than 2 per cent of total observations, therefore the impact is marginal. 


\subsection{Flare detection}\label{flaredetection}
As the SSO data sample is constrained to only 154 targets, it is possible to identify flares manually. Therefore we did not find it necessary to implement a fully automated, complex flare-detection algorithm. Instead we decided to perform flare detection in two parts: a simple, automated flare-detection algorithm to extract all the flare candidates, followed by a manual vetting process to confirm them. Both stages of this flare detection process are demonstrated in Figure \ref{fig:flare} for target Gaia DR2 3200303384927512960. We note that this target is one of the most frequently flaring objects studied by \citet{Seli2021ActivityTESS}. This object is a particularly challenging case due to its rapid variability, where flare decays are difficult to separate from photometric modulations.

\subsubsection{Automatically searching for flares using lightcurve gradients}\label{automaticflaredetection}
For the first part of our flare detection process, we chose to implement the automatic flare detection method set out in  \cite{Lienhard2020GlobalSurvey}, due to its simplicity, speed, and robustness. This method was developed on lightcurves from the TRAPPIST (TRAnsiting Planets and PlanetesImals Small Telescope) telescopes \citep{Gillon2011TRAPPIST:Systems,Jehin2011TRAPPIST:Telescope}. \cite{Lienhard2020GlobalSurvey} capitalise on the changing gradients in the asymmetric structure of a flare, which involves a sharp increase to a peak followed by a slower exponential decay \citep{Moffett1974UVDATA}. They evaluate the following two criteria:
\begin{equation}\label{eq:florianflare1}
    \frac{2f_{j}-f_{j-2}-f_{j+3}}{\sigma_{j}}\times
    \frac{|f_{j}-f_{j-2}|-|f_{j+1}-f_{j}|}{\sigma_{j}} > A_{\textrm{thresh}} 
\end{equation}
and
\begin{equation}\label{eq:florianflare2}
    2f_{j} - f_{j - 2} - f_{j+3} > 0,
\end{equation}
where $f_{j}$ is the flux of the $j^{\textrm{th}}$ data point in the target's lightcurve, and $\sigma_{j}$ is the RMS of the section of the lightcurve comprised of the nearest 60 data points. By only considering small neighbouring sections of the lightcurve in our RMS, and the flux for a few points either side of the flare in our criteria, we are able to detect small flares on rapidly rotating or frequently flaring objects. Equation \ref{eq:florianflare1} assesses the quality of the flare's shape. If we assume that the peak of the flare is at $j$ then the first half of Equation \ref{eq:florianflare1} confirms that we are at a peak by ensuring the flux at $j$ is greater than the flux a few exposures before and after. The steeper the peak, or larger the flux difference between the peak and surrounding points, the higher its value. The second half of Equation \ref{eq:florianflare1} hinges on the asymmetry of the flare by requiring that the gradient before the peak, the fast rise, is larger than the gradient after the peak, the slow decay. For this second half of the equation, the greater the asymmetry, the higher its value. By dividing by the local RMS, the aim is to remove flare structures caused by photometric scatter, which is especially problematic for ground-based observing with rapidly changing weather conditions. Equation \ref{eq:florianflare2} simply prevents the case where both halves of Equation \ref{eq:florianflare1} are negative. 

It is possible to have significant time gaps within the lightcurve sections of 60 points due to bad weather, however, this scenario is rare. Including data points that are far apart in time would lower the reliability of the RMS in this section of the lightcurve, however, when weather conditions are adverse enough to close the telescope, or seriously affect photometric precision, the RMS directly before or after is likely to be enhanced (due to poor observing conditions that do not quite meet the threshold to be removed). We note that the ``nearest 60 points'' are only used for this flare detection method, following \citet{Lienhard2020GlobalSurvey}, whereas time boxes are used for calculating the running median and RMS in Sections \ref{smalllargeflares}, \ref{validate} and the rest of this work.

In their paper, \cite{Lienhard2020GlobalSurvey} determine $A_{\textrm{thresh}}=12$ by inspection of the smallest flares that they intended to remove; however, this value is specific for the photometric precision and typical exposure times of TRAPPIST lightcurves. We would expect the precisions obtained by TRAPPIST's 60-cm telescopes to differ significantly from those achieved by SPECULOOS's 1-m telescopes. Instead, for the SSO lightcurves, we derived a value of $A_{\textrm{thresh}}=5$. We find a lower value for $A_{\textrm{thresh}}$ partly due to the increased photometric precision of the SSO and partly because we are choosing to manually verify the flare candidates that this algorithm detects, so we can afford to over-detect at this stage.  

By choosing a flare detection method based on the shape of a flare, and not outlier detection, this may limit the diversity of flare morphologies that we will detect. Since there have been few large-scale studies done to date on ultra-cool dwarf flares, such as those in our SSO sample, any difference in flare structure between earlier- and later-spectral-type objects is still unknown. Therefore, we decided to focus on the flares most resembling the standard flare shape \citep{Moffett1974UVDATA}, while flagging more unusual lightcurve behaviour in the manual vetting stage (see Section \ref{manualvetting}).

\subsubsection{Small vs. large flares}\label{smalllargeflares}
To both maximise the detection of small flares, and optimise the modelling of high-amplitude flares with slow recovery times, we split the flare candidate sample into two categories: \textit{small} and \textit{large} flares. We classify \textit{large} flare candidates as those with at least 2 data points in the flare region more than 7 times the running local standard deviation (standard deviation of the surrounding 80 data points) above the running local median (defined similarly as the median of the surrounding 80 data points); otherwise, we classified it a \textit{small} flare candidate. The flare region starts at least two points before the peak. We approximate the end of the flare region by using least-squares to fit the flare decline with a sum of a fast and a slow-decaying exponential, as in \citet{Davenport2014Kepler1243}. The decay time of the slower exponential decay is used to estimate the time of the end of the flare region.

\subsubsection{Validating and vetting flares}\label{validate}
Once a flare candidate is classified as \textit{large}, then we use the surrounding 320 points (defined as global), instead of 80 points (local), to calculate the running median and standard deviation for the next step of validation.

If the criteria in Section \ref{smalllargeflares} is met, we then run several additional automatic quality checks to separate flare candidates from flare-like signals replicated by noise. By ensuring that there is more than one data point in the flare region after the peak, we remove cosmic rays from the flare candidates. Additionally at least 2 points in the flare region must be more than 2 times the running local standard deviation above the running local median of the lightcurve. The local standard deviation and median are as defined in Section \ref{smalllargeflares}. When calculating the running local median and standard deviation, we mask all potential flare candidates in the lightcurve, and then interpolate over the flare region.

This automated algorithm provides us with a collection of flare candidates to follow up with manual vetting to then obtain our final flare sample.

\subsubsection{Manual vetting}\label{manualvetting}
Once we have a collection of flare candidates from the method detailed in Sections \ref{automaticflaredetection} and \ref{validate}, then the second part of our flare detection process is to manually inspect these candidates and obtain our final flare sample. We consider each flare candidate and only confirm those which can be clearly identified as matching the standard flare shape of a sharp flux increase followed by a slower, exponential flux decay \citep{Moffett1974UVDATA,Davenport2014Kepler1243}. We also ensure that the flares cannot be attributed to rapid changes in atmospheric conditions. 

As a final validation step, we check through the global lightcurves in the SSO sample. The only flare-like structures remaining are due to the following events: a flare occurs before the start of the night (where we catch the tail end of the flare) or at the very end of a night, the structures are low-amplitude and so cannot be distinguished from noise, the structure diverges significantly from the standard flare model, or they are correlated with systematics. Examples of structures identified in the SSO data set that deviate from the standard flare model, and are not clearly overlapping flares, are symmetric flares (with similar rise and decay times) and flares with a sudden increase and linear decline. During this manual vetting stage we do not add any of these potential missed flares into the flare sample, as they do not meet the criteria to be clearly identified as a flare with the standard flare shape.

\begin{figure*}
    \centering
    \includegraphics[width=\textwidth]{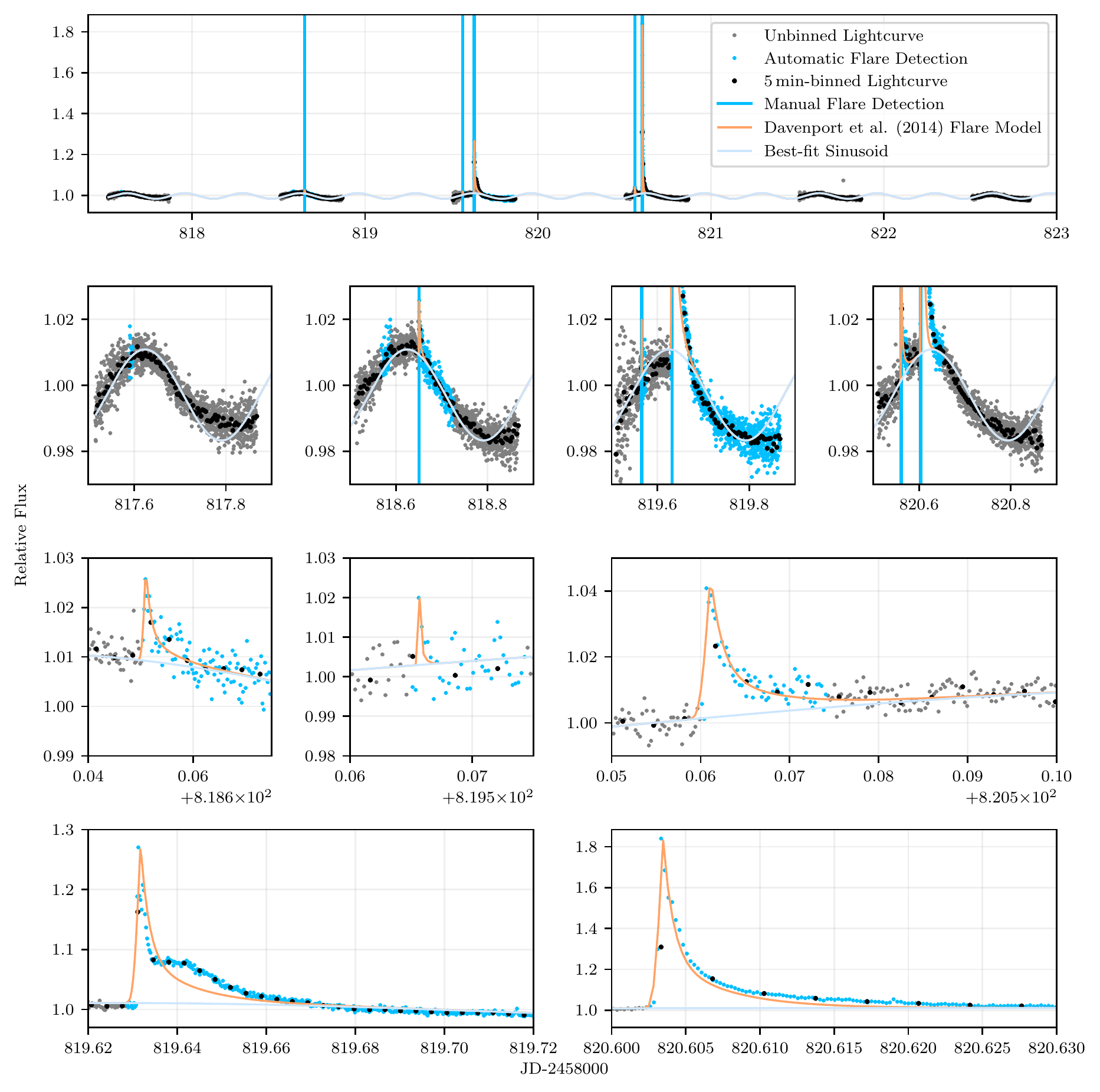}
    \caption{Demonstration of the flare detection process on six nights of observations of an M7 object ($J$ = 10.7 mag), which exhibits both frequent flaring and short-period rotation. The four rows show different views of the same global lightcurve, zoomed in to individual nights on the second row, and only the five manually-verified flares on the bottom two rows. The grey points are the unbinned data, while the black points are binned every 5 minutes. We do a simple least-squares fit of a sine wave with a period of 0.33d, shown in light blue. It is clear from this fit that the rotation pattern is not perfectly sinusoidal; however, it provides a way to visualise the periodicity. The medium-blue data points are those initial flare regions flagged by the automatic flare detection algorithm described in Section \ref{automaticflaredetection}. The vertical, medium-blue lines are the flare candidates that have also been confirmed manually. During visual inspection of this lightcurve, we remove two flares (one in the first plot on the second row and the first flare in the second plot on the second row), due to their small amplitude, which were found too difficult to detach from the photometric scatter. The best-fit \protect\citet{Davenport2014Kepler1243} models for each flare are shown in orange. The case where we have a poor fit to the template is shown in the bottom left plot.}
    \label{fig:flare}
\end{figure*}

\subsection{Modelling flares}\label{modelflares}
In order to model each flare, we coarsely flatten the lightcurve surrounding that flare by dividing the lightcurve by a median filter, using the local or global criteria as in Section \ref{smalllargeflares}.

We chose to model our flares by fitting the empirical flare template described in \cite{Davenport2014Kepler1243}. In this paper the authors generate a median flare template from Kepler observations of 885 `classical' flares on the M4 star GJ 1243. They found a sharp flux increase, modelled with a fourth-order polynomial, followed by an initial fast exponential decay and subsequent slower exponential decay. This model requires a time for the peak of the flare, a relative amplitude on the normalised lightcurve, and a full width half maximum (FWHM$_{\textrm{flare}}$), which corresponds to the flare's decay time. We demonstrate fitting this flare model to flares of differing amplitudes and decay times in Figure \ref{fig:flare}.

While this model works well for the more classically shaped flares, it struggles to represent complex flares that do not fit the standard flare morphology \citep{Davenport2014Kepler1243}. This template has yet to be tested on a statistically large number of low-mass objects, such as UCDs, whose flare profiles may differ significantly. Complex flares include those with multiple peaks \citep{Davenport2014Kepler1243}, oscillations \citep{Anfinogentov2013THEMEGAFLARE}, and those which are closely entangled with variability. The overlapping of multiple flares is likely the culprit of more unusual lightcurve structure. During this manual vetting, we divide flare regions containing clear multiple peaks into separate flares and each of the new separated flares are added to the flare list. We also flag (but do not remove) 27 flares that either do not fit the standard flare model, or are not easily separable into multiple flares. This corresponds to 11 per cent of the total flare sample. An example of a possible flare overlap is shown in the bottom left plot in Figure \ref{fig:flare}.

\subsection{Extracting flare energies}\label{flareenergies}
To measure the flare energies, we follow the technique described in \cite{Shibayama2013SuperflaresSuperflares}. We model the flare as blackbody with an effective temperature, $T_{\textrm{flare}}$, of 9000$\pm$500\,K \citep{Kowalski2013Time-resolvedSpectra}, and we assume this temperature remains constant. We calculate the luminosity of each star and flare, as seen through the SSO's I+z' filter, accounting for the overall system efficiency, as follows:
\begin{align}\label{eq:lstar}
    L'_{*} &= \pi R_{*}^2\,\int_{I+z'}{R_{\lambda} B_{\lambda}(T_{\textrm{eff}}) ~\textrm{d}\lambda} \\
    L'_{\textrm{flare}} &= A_{\textrm{flare}}(t)\int_{I+z'}{R_{\lambda} B_{\lambda}(T_{\textrm{flare}}) ~\textrm{d}\lambda}. 
\end{align}
Here, $R_{\lambda}$ is the total SSO response function, which is the product of the transmission in the $I+z'$ filter, quantum efficiency of the CCD, CCD window, and reflectivity of the mirror coatings. $B_{\lambda}(T_{\textrm{eff}})$ and $B_{\lambda}(T_{\textrm{flare}})$ are the Planck functions evaluated for the star’s effective temperature, $T_{\textrm{eff}}$, and the flare temperature, respectively. Finally, $A_{\textrm{flare}}(t)$ is the area of the flare. We do not include atmospheric transmission in our response function due to its rapid variability. Since the relative flare amplitude can be found directly from the normalised lightcurve, $F(t) = (\Delta F/F_{\textrm{mean}})(t) = L'_{\textrm{flare}}(t)/L'_{*}$, we can solve for $A_{\textrm{flare}}$ as follows:
\begin{equation} \label{eq:aflare}
    A_{\textrm{flare}}(t) = F(t)\,\pi R_{*}^2\,
    \frac{\int_{I+z'}{R_{\lambda} B_{\lambda}(T_{\textrm{eff}}) ~\textrm{d}\lambda}}{\int_{I+z'}{R_{\lambda} B_{\lambda}(T_{\textrm{flare}}) ~\textrm{d}\lambda} }.
\end{equation}
To best estimate the flare amplitude, we use the local or global running median for $F_{\textrm{mean}}$, as defined at the end of Section \ref{smalllargeflares}. The bolometric flare luminosity, $L_{\textrm{flare}}$, can then be calculated with the assumption that the star radiates as a blackbody:
\begin{equation} \label{eq:lbol}
    L_{\textrm{flare}} = \sigma_{\textrm{SB}} T_{\textrm{flare}}^4 A_{\textrm{flare}},
\end{equation}
where $\sigma_{\textrm{SB}}$ is the Stefan-Boltzmann constant. The total bolometric energy of a flare ($E$) is then the integral of $L_{\textrm{flare}}$ over the flare duration:
\begin{equation} \label{eq:eflare}
    E=\int_{\textrm{flare}}{L_{\textrm{flare}}~\textrm{d}t}.
\end{equation}
Using Equation \ref{eq:lbol}, and assuming the flare temperature is constant, we can separate the time-dependent and independent components:
\begin{equation} \label{eq:flaressep}
    E=C\int_{\textrm{flare}}{A_{\textrm{flare}}~\textrm{d}t},
\end{equation}
where $C=\sigma_{\textrm{SB}} T_{\textrm{flare}}^4$. Using Equation \ref{eq:aflare} we can further separate out the time-independent components that depend on the properties of the star and the flare, and the time-dependent integral:
\begin{equation} \label{eq:flaresimp}
    E=C_{\textrm{star}}(T_{\textrm{eff}},R_{*})C_{\textrm{flare}}(T_{\textrm{flare}})\int_{\textrm{flare}}{F(t)~\textrm{d}t}.
\end{equation} 
The motivation for calculating the energy in this way is that it allows us to calculate these three components separately. When generating a large pool of synthetic flares for injection-recovery tests, as in the following section, this drastically reduces the processing time. Every object in our SSO data sample will have a constant value for $C_{\textrm{star}}$, which only needs to be calculated once per star. The value for $C_{\textrm{flare}}$ also becomes constant with our assumptions. It is then only the integral of the flare in the normalised lightcurve that remains to be calculated. 

For this integral we decided to use the normalised lightcurve (corrected with the same local or global median filter described in Section \ref{modelflares}) directly for the energy calculation, rather than using the flare model. In doing so we note that the energies we calculate provide a lower bound on the flare energy, as we may miss the flare's peak. It is also possible that if there was a large flare on a rapidly-rotating star, the underlying $F_{\textrm{mean}}$ may `smooth out' photometric structure, leading to the flare energy being over-estimated. However, to be classed a `large' flare, a flare needs to be more than 7 times the running global standard deviation (as defined in Section \ref{smalllargeflares}). This standard deviation would be highly inflated from the periodic photometric modulations, and therefore this flare would have to be very high energy for this to become an issue. As high energy flares are rare \citep{Gershberg1972Some1967-71, Lacy1976UVData.}, this scenario is not a major concern. We chose not to use the flare model because the flares in our lightcurves are well-sampled, and when there is increased photometric noise in the lightcurve, or more complex flare shapes from overlapping flares (see Section \ref{modelflares}), the fit to the flare model can be unreliable. Integrating the lightcurve directly means that bursts of flares occurring in short succession are counted as one flare of increased energy. While this simplification may affect the calculation of our flare rates and energies, it should not significantly affect our results as we only detect 27 unusually-shaped flares that may be a blend of multiple flares. We note that while flares are typically approximated as blackbodies with temperatures from 9000--10000\,K, the flare temperature has been observed to vary outside this range, both between and within flare events \citep{Howard2020EvryFlare.TESS}.

\subsection{Calculating Flare Rates}\label{flarerates}
As flares are stochastic events, the gaps in observations due to the day-night cycle or bad weather loss do not hinder our ability to calculate flaring rates, which are only dependent on the total time on-sky. It is likely that we underestimate flaring rates for the several reasons flares are not found by our automatic detection algorithm, outlined in Section \ref{manualvetting}.

We calculate our flare rates as the number of flares detected per target divided by the observation time (summed across all telescopes). Almost all of our targets have been observed for less than 200 hours, therefore we have a lower limit on individual stars' flaring rates of $\sim$0.12\,$\textrm{d}^{-1}$.

\subsection{Completeness of the flare sample}\label{injectionrecovery}
To evaluate the completeness of our flare sample (the minimum energy flares recoverable by our flare detection process for different spectral types), we perform artificial flare injection-recovery tests. Using the global lightcurves from the 154 different targets with at least 20 hours of observation, we mask all of the flares detected in Section \ref{flaredetection}. We use the \cite{Davenport2014Kepler1243} flare model to generate 100,000 artificial flares with amplitudes drawn from a log-normal distribution between relative fluxes of 0.001 and 5, and FWHM$_{\textrm{flare}}$ drawn from a uniform distribution between 30 seconds (typical exposure time for SPECULOOS) and 1 hour. We calculate the energy of the resulting flares for each star, and divide into 6 energy bins in log$_{10}$ space from $10^{28}$ to $10^{34}$\,erg. 

We note that here we have followed the work of \citet{Davenport2014Kepler1243,Davenport2016THEFLARES,Gunther2020Dwarfs}, among others, and decided not to consider any relationship between amplitude and duration. While we may be injecting ``unphysical'' flares, this approach allows us to explore a larger parameter space of different flare morphologies. The amplitude-duration relationship for flare has also not been well studied for the case of UCDs, where it may differ from that seen in earlier M dwarfs. We do not see a clear relationship between fitted amplitude and duration in our sample (and so cannot confirm any other observed amplitude-duration relationships, e.g. \citealt{Hawley2014KeplerDwarfs}), though our estimates of amplitude and duration rely on a good fit to the Davenport template, which is significantly more challenging with messy ground-based data.
Our purpose with these tests is to examine the decay time, amplitude, energy and spectral type limits of our flare detection method, not to reproduce a realistic flare frequency distribution. To model a realistic flare distribution we would need to consider that low energy flares are more common than high energy flares, as discussed in Section \ref{ffd}. However, then we would have significantly more low energy flares injected and recovered, making our sensitivity results more robust for small amplitudes/decay times, and less robust for high energy flares.

For each energy bin, treated independently, we then randomly select 5 flares and inject them into our differential lightcurves at a random observed time, taking care to ensure the $\pm$5 data points surrounding the time of the flare's peak are within 0.01\,d (14.4\,mins). This prevents those 5 flares from occurring too close to the start or end of night or during gaps in our observations, which we would always miss. We allow flares to overlap to reflect the scenario we see often in real lightcurves, however as there are only ever a maximum of 5 flares in a lightcurve at once, the impact on our recovery results is limited.

We then run the automatic part of our flare detection method, described in Sections \ref{automaticflaredetection} and \ref{validate}, on all injected lightcurves and record the recovered flares. We consider a flare to be recovered if the recovered time for its peak flux is within 1 FWHM$_{\textrm{flare}}$ (of that flare) of the injection time. We allow for a more flexible recovery time because of the decision to allow flares to overlap, which can lead to more complex structures with multiple peaks. If two or more flares overlap within 1 FWHM$_{\textrm{flare}}$, they may be detected as only one flare. We assume if they occur further apart in time then they should be easily separable. This process is then repeated for each energy bin 5 times, making sure to use different flares from our artificial sample in each iteration.  

The manual vetting step of our flare detection technique was not feasible for checking the approximately 23,000 flares we injected. This may introduce a bias into our results. It is more likely during manual vetting that we would conflate low-amplitude flares or flares on high photometric scatter targets with noise, therefore by removing this stage we have potentially inflated our recovery rates for the lowest energy flares, or for the faintest target stars.

\begin{figure}
	\includegraphics[width=\columnwidth]{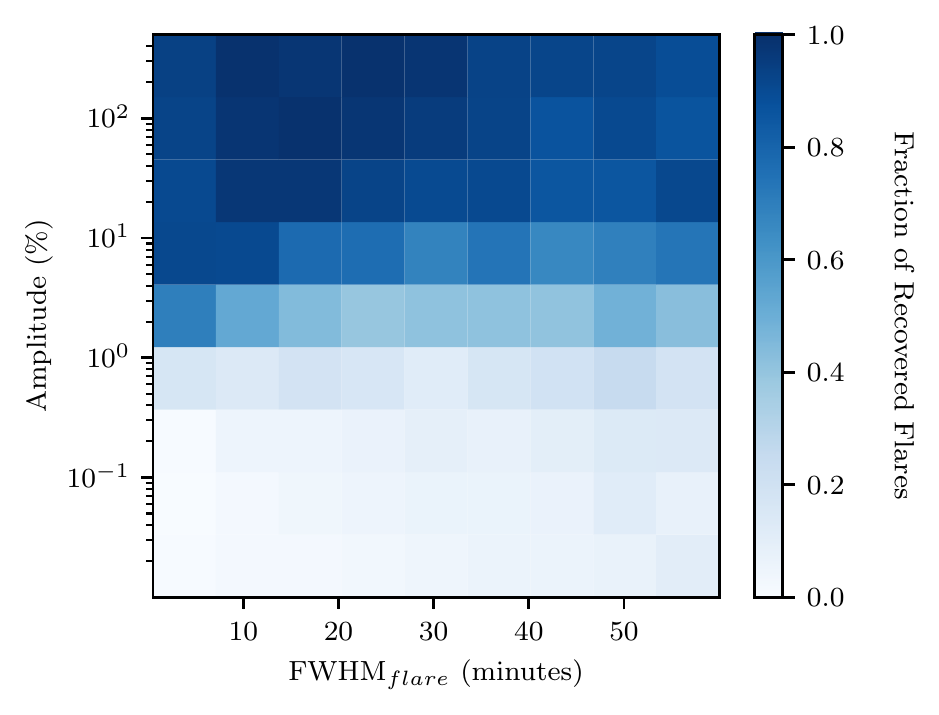}
    \caption{Fraction of injected flares recovered as a function of flare amplitude and FWHM$_{\textrm{flare}}$ using the automatic flare detection algorithm described in Section \ref{flaredetection}.}
    \label{fig:inject_recover}
\end{figure}

Figure \ref{fig:inject_recover} shows the results of the injection-recovery tests, carried out for approximately 23,000 artificial flares. We are able to detect a significant proportion of flares with amplitudes above 1 per cent. Due to the short exposure times of SPECULOOS, we have the advantage of being very sensitive to flares with a short duration (with FWHM$_{\textrm{flare}}<5$\,minutes). However, we are more limited in detecting longer duration flares (with FWHM$_{\textrm{flare}} \geq60$\,minutes) due to the day-night cycle, with typical uninterrupted observation windows of 4--8 hours. There will also be some stars which have nights of observation less than 4 hours, due to weather, or limited visibility, and these will be the most difficult targets on which to detect slowly decaying flares. However, we note that during the manual vetting (Section \ref{manualvetting}) we inspect the lightcurves with flares removed and we do not find any undetected high energy flares remaining in our sample. We also assess our limitations with the photometric RMS of our lightcurves. When the RMS of a global lightcurve exceeds $\sim0.7$ per cent (for 5-minute binning) we begin to see a drop in detection efficiency, resulting in an increase in the minimum detectable amplitude. Only three targets in our dataset exceed this RMS.

\begin{figure}
	\includegraphics[width=\columnwidth]{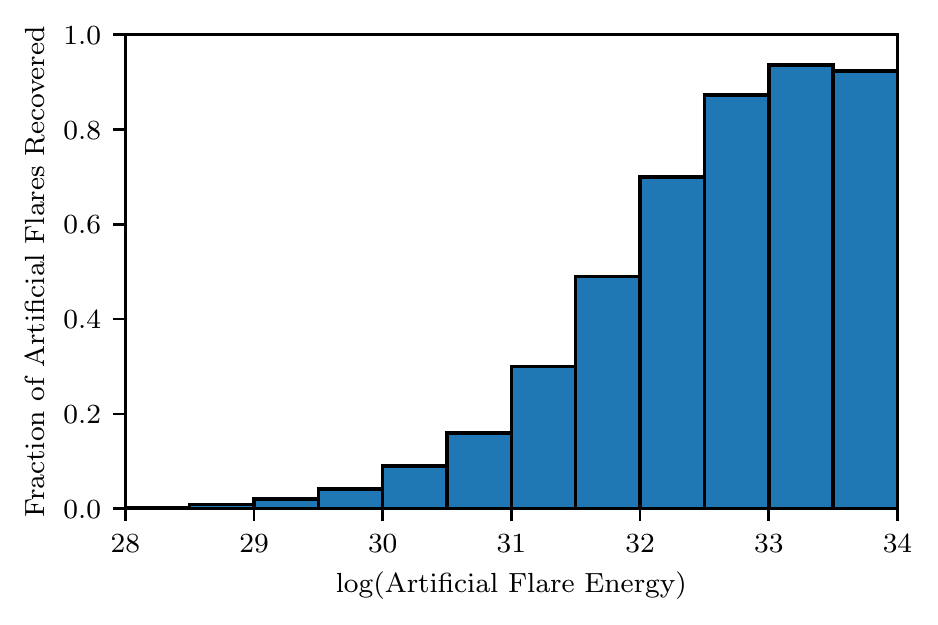}
    \caption{Fraction of injected flares recovered as a function of flare energy.}
    \label{fig:recover_energies}
\end{figure}

\begin{figure}
	\includegraphics[width=\columnwidth]{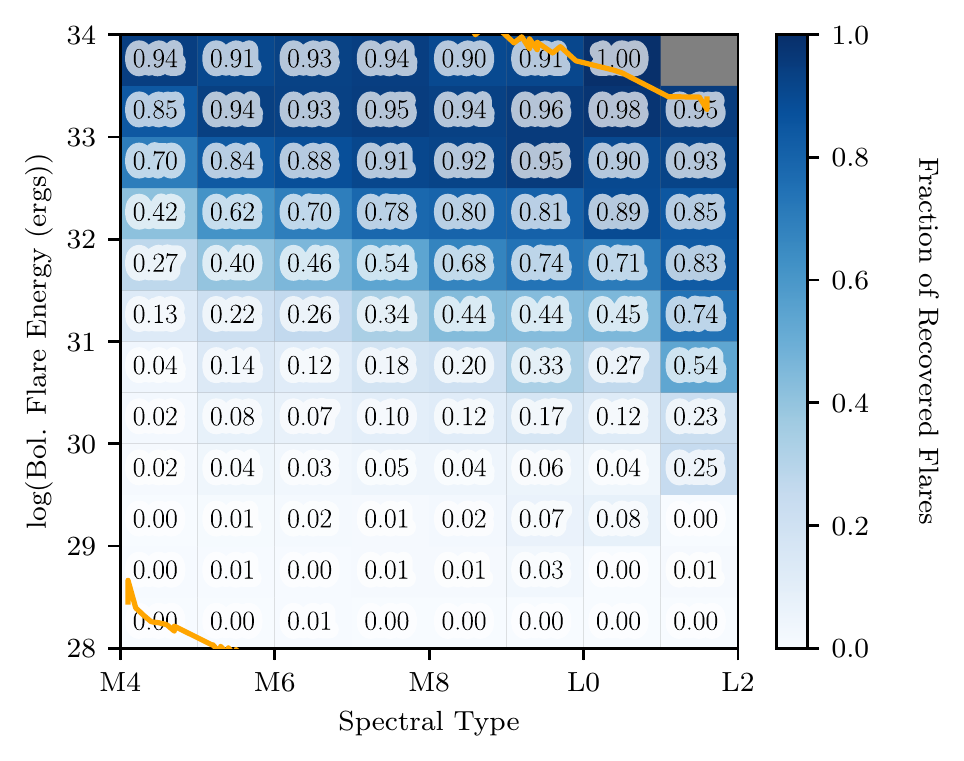}
    \caption{Fraction of injected flares recovered as a function of flare energy and the object's spectral type. We find that as we move to cooler stars we are able to detect lower energy flares. The upper and lower flare energy limits, determined by the amplitude and $\textrm{FWHM}_{\textrm{flare}}$ limits of the artificial flare sample and the radius and $\textrm{T}_{\textrm{eff}}$ of each star, are shown in orange. We are only impacted by the sample limits for spectral types earlier than M5 and later than L0.}
    \label{fig:recover_energies_teff}
\end{figure}

For a given flare amplitude and FWHM$_{\textrm{flare}}$, the flare energy will vary depending on the effective temperature and radius of the star (Equation \ref{eq:flaresimp}). The recovery fraction for different flare energies across all stars is shown in Figure \ref{fig:recover_energies}. However, the minimum detectable energy of recovered flares varies with spectral type, as shown in Figure \ref{fig:recover_energies_teff}, which facilitates the detection of lower energy flares on our coolest dwarfs. For low-energy flares, we have two competing biases. The lightcurves for the later, and fainter, M and L dwarfs have higher photometric scatter, which makes it difficult to detect small-amplitude (and therefore lower energy) flares. However, as flares have a strong white light component \citep{Namekata2017StatisticalStars}, there is an increased contrast between flares and the stellar spectra of red dwarfs, which should make lower energy flares easier to identify in our coolest stars \citep{Allred2015AFlares, Schmidt2019TheASAS-SN}. SPECULOOS is also optimised to observe ultra-cool objects, and due to this specificity, we only see a very minimal increase in RMS with later spectral types.

The results from our artificial flare injection-recovery tests allow us to calculate the recovery fraction of our flare detection method, $R(E)$, that reflects the low recovery rates of low energy flares. The average $R(E)$ for all stars is shown in Figure \ref{fig:recover_energies}, as well as its spectral dependence in Figure \ref{fig:recover_energies_teff}. As our sample is already small we are unable to reduce detection-sensitivity effects by simply limiting our sample to flares with energies above a minimum recoverable energy \citep{Paudel2018Dwarfs}, or energies where a sufficient fraction of flares are recovered \citep{Davenport2016THEFLARES}. Instead, we decided to calculate flare occurrence rates in Section \ref{senstivitymap} using the method in \citet{Jackman2021StellarSurvey}, which models the decline in sensitivity for low energy flares.


\section{Identifying rotation periods}\label{rotation}
With the previously identified flares and bad observations masked, we search for rotation periods in the 154 low-mass objects with more than 20 hours of observation. We apply the Lomb-Scargle periodogram analysis \citep{Lomb1976Least-SquaresData, Scargle1982StudiesData} to our global lightcurves, binned every 20 minutes. The SSO lightcurves are not uniformly sampled, as we have gaps in our data, not only from the day/night cycle, but also from bad weather, masked flares, and changes to our observation strategy. Therefore, careful treatment of the resulting periodograms is essential to remove aliases.

We searched for periods between the Nyquist limit (twice the bin size) and the entire observation window. While we cannot be fully confident in periods greater than half their observation window, by expanding the period range we search, we can estimate a lower bound on long-period rotation. We then remove all peaks in the periodogram with a false alarm probability below 3$\sigma$ (0.0027). Any target which has at least one significant Lomb-Scargle peak undergoes visual inspection of their periodogram and the corresponding phase-folded lightcurves for all possible rotation periods (all significant peaks in the periodogram). In addition, by comparing with periodograms of the time stamps, airmass, and FWHM, we can eliminate signals which arise from the non-uniform sampling or ground-based systematics.

We decided to apply a similar classification system as \cite{Newton2016RotationProject} to our rotating objects. Classifying our rotators helps account for the difficulties in observing from the ground, both through atmospheric systematics, and through irregular, non-uniform sampling. When examining each lightcurve, we ask ourselves several questions:
\begin{enumerate}
    \item Is the period clearly visible in the phase-folded lightcurve?
    \item Was the object observed for long enough to span multiple periods?
    \item Is the frequency an alias of the ``day signal", seen as integer multiples in frequency space (periods of 1, 0.5, 0.33\,d etc.)?
    \item Is there a correlation with systematics (as seen in the airmass or FWHM periodograms)?
    \item Can the period be seen by eye in the un-phased lightcurve?
    \item If we observed this object with more than one telescope, does this period fit them all?
    \item Can we easily disentangle the `real' period from its one day aliases?
    \item Is the amplitude of the periodic signal above the level of noise in the lightcurve?
\end{enumerate}

If a rotator passes all the above criteria, then we class it as `A'. If it fails any of the above, but the rotation still seems likely then we class it as a `B' grade rotator. Most commonly, the `B' rotators are convincing, but we do not observe multiple cycles, or we cannot easily choose between the period and its one-day aliases. Any lightcurves for which we see some periodic structure, but we cannot easily determine a period, we class as `U'. This can result from broad peaks in the Lomb-Scargle periodogram, or multiple possible periods due to lack of observations, or a very low amplitude periodic signal that is comparable to the noise level. Because of these ambiguities in the period measurements, we do not attempt to place errors on our period estimates. If we do not detect any periodic signal or we cannot remove correlations with systematics (such as for very crowded fields), then we class the object as `N'. Finally, in addition to the classes defined in \cite{Newton2016RotationProject}, we add an extra `L' grade. This is for the lightcurves where we see clear long-period rotation, but the period is similar or longer than the time window observed. For these objects the best we can do is to estimate the lowest possible period we measure, and acknowledge that there are large uncertainties on these period values. 


\section{Results}\label{results}
The SSO data sample, along with the results of the flare and rotation analyses, are presented as supplementary material. 

\subsection{SPECULOOS-South Flare Sample}
\begin{figure}
    \includegraphics[width=\columnwidth]{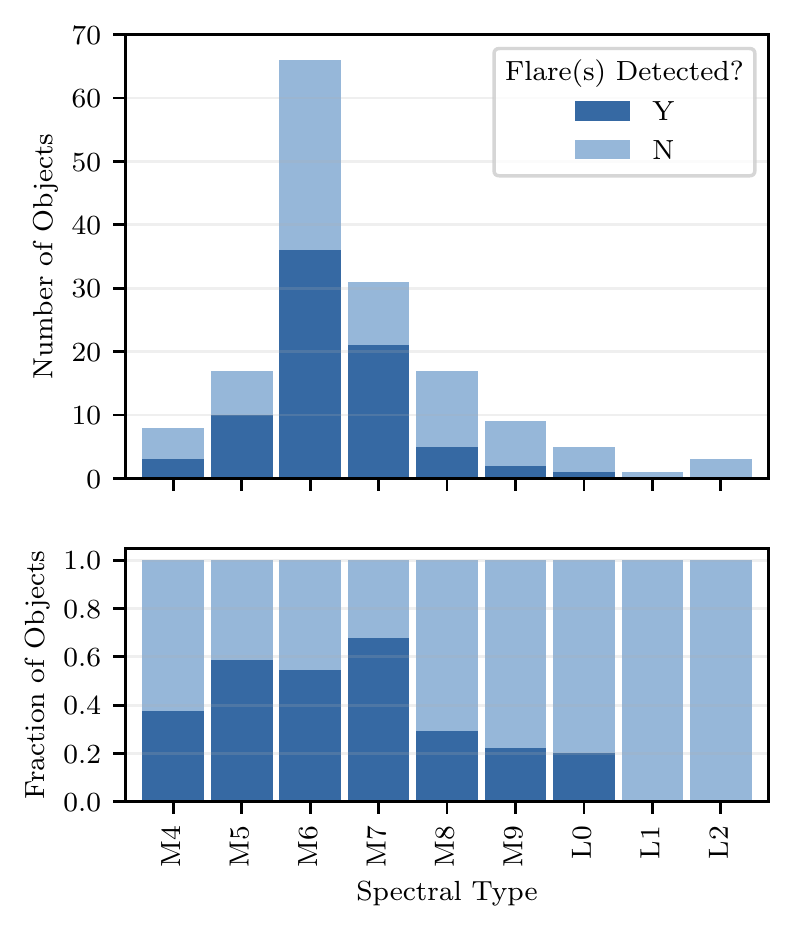}
     \caption{\textbf{Top:} Histogram of objects in the SSO data sample as a function of spectral type. The flaring objects are marked in light blue. \textbf{Bottom:} Fraction of flaring objects as a function of spectral type. }
     \label{fig:flares_spec}
\end{figure}

From the SSO dataset described in Section \ref{ssosample}, we identify 234 flares. We find that of our 154 unique targets, 78 are flaring (50 per cent). These flaring stars span the spectral type range from M4 ($\textrm{T}_{\textrm{eff}}=3160\textrm{K}$) to L0 ($\textrm{T}_{\textrm{eff}}=2313\textrm{K}$). Figure \ref{fig:flares_spec} shows the spectral type distribution and proportions of the flaring objects identified in the SSO sample. From this figure we see the proportion of flaring stars stays consistently above 60 per cent for objects of spectral type M5--M7. We see that the rate of flaring stars begins to decline around M8 ($\sim30$ per cent), with no detected flares for any object beyond L0. The coolest flaring star we detect is a 2313\,K, M9.6V object (which is rounded to L0 in Figure \ref{fig:flares_spec}). However, for the L dwarfs we are limited by the small sample size, and therefore cannot make any conclusions about whether the fraction of flaring objects continues to decrease beyond late M dwarfs. Likewise, we do not have any objects earlier than M4; however, these earlier M dwarfs have been well studied by TESS, Kepler, and MEarth. Despite our small sample of M4 dwarfs, we see an apparent reduction in activity for stars earlier than M5 that agrees with the previous results ($\sim$30 per cent for M4 with TESS, \citet{Gunther2020Dwarfs}, 25\% for M5 with ASAS-SN, \citet{Martinez2019AASAS-SN}). Several authors find a steep rise in flaring fractions around spectral type M4 \citep{West2004SPECTROSCOPICSUBDWARFS, Yang2017TheField, Gunther2020Dwarfs, Martinez2019AASAS-SN}. Specifically, \citet{West2004SPECTROSCOPICSUBDWARFS} analysed the H$\alpha$ emission of cool stars (which roughly correlates with flaring activity, \citealt{Yang2017TheField, Martinez2019AASAS-SN}) and determined that the fraction of active stars rose monotonically from spectral type M0 to M8, peaked at M8 ($\sim$80 per cent active) and then declined to L4. Our results show an increase from M4 to a broad peak from M5--M7, followed by a decrease from M8 to L2.

We probe the parameter space of high-flare-rate stars with low-to-mid energy flares. Due to the relatively short baselines of SPECULOOS observations, if a target does not flare frequently, then it is unlikely we would identify it as flaring. This also means we are less likely to detect the rarer, high-energy flares \citep{Gershberg1972Some1967-71, Lacy1976UVData.}. In conjunction with the difficulties arising from the day-night cycle (typical night observations are 4--8 hours), which complicates our detection of very slowly decaying flares (see Figure \ref{fig:inject_recover}), it is unlikely that superflares with energies $>10^{33}$\,erg \citep{Shibayama2013SuperflaresSuperflares} will be detected by our survey. From studying the completeness of our flare sample, we also conclude it is unlikely that we will detect any flares below $10^{29.5}$\,erg for any spectral type. 

\subsection{Flare energies}
In this flare sample, we recover flares ranging in energy from $10^{29.2}$ to $10^{32.7}$\,erg, with a median of $10^{30.6}$\,erg. We do not detect any superflares, with energies between $10^{33}$--$10^{38}$\,erg \citep{Shibayama2013SuperflaresSuperflares}.

From our flare injection-recovery analysis we see that we have a spectral dependence in the flares we can retrieve. This dependence means that we will struggle to recover flares on the earliest M dwarfs below an energy of $\sim$10$^{31}$\,erg. As we move towards later spectral type objects, we are able to recover significantly higher fractions of lower energy flares, as shown in Figure \ref{fig:recover_energies_teff}. 

\subsection{Flare frequency distribution}\label{ffd}
Flare Frequency Distributions (FFDs) explore how often a star will flare with at least a certain energy. The FFD assumes that the following power law applies \citep{Gershberg1972Some1967-71,Lacy1976UVData.,Hawley2014KeplerDwarfs}):
\begin{equation} \label{eq:power_law2}
    dN(E)=kE^{-\alpha}dE,
\end{equation}
where N is the flare occurrence rate, $E$ is the flare energy, and k and $\alpha$ are constants. This can also be represented as:
\begin{equation}\label{eq:powerlaw}
    \log(\nu) = C + \beta \log(E),
\end{equation}
where $\nu$ is the cumulative frequency of flares of energies $\geq E$, $C=\log(\frac{k}{1-\alpha})$ and $\beta=1-\alpha$. We linearly fit for the two coefficients, $C$ and $\beta$, using a least-squares optimisation. Physically, where $\beta>1$ low-energy flares have the largest contribution to the total energy emitted by flares, whereas when $\beta<1$, it is the highest energy flares that make up the majority of this total energy.

\begin{figure}
	\includegraphics[width=\columnwidth]{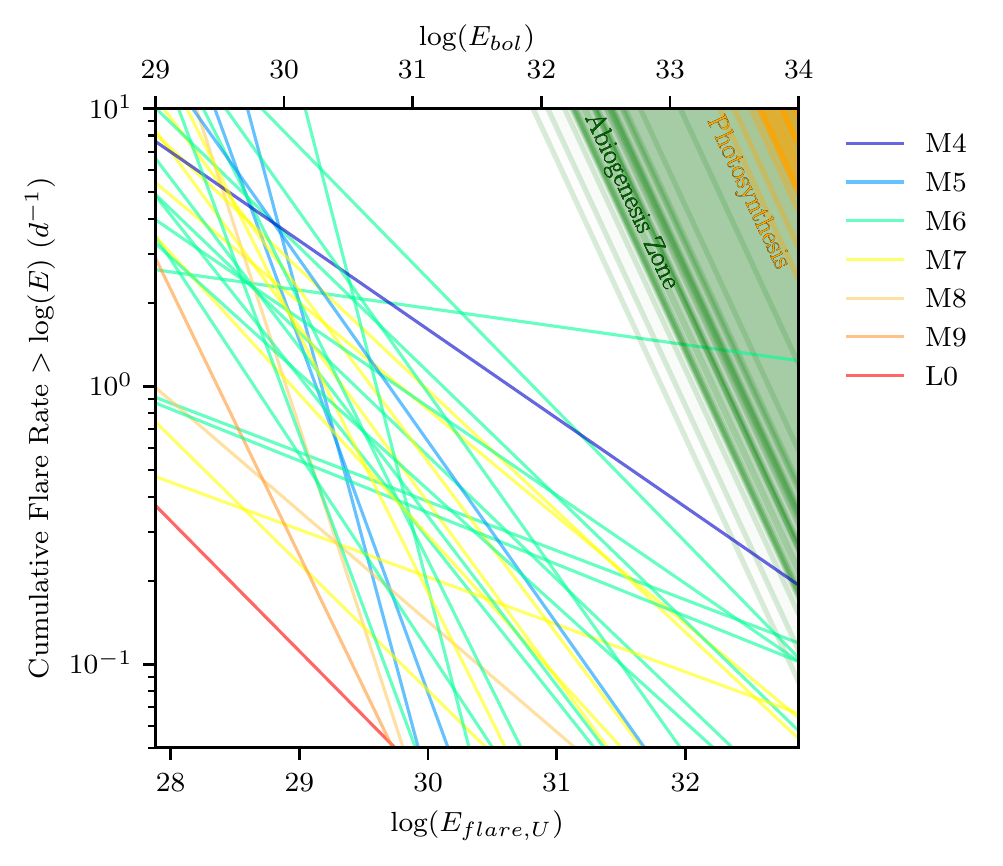}
    \caption{Flare frequency distributions for every star with at least 3 flares. We find the best power-law fit to the cumulative flare rates against flare energy for every star, shown by straight lines. The spectral type of each star is indicated by its colour. The abiogenesis zones \citep{Rimmer2018TheExoplanets} are also shown for each star as the regions shaded in green (similarly to \citealt{Gunther2020Dwarfs}), and the photosynthesis thresholds \citep{Lingam2019PhotosynthesisStars} are shown in orange. }
    \label{fig:ffd}
\end{figure}

To generate FFDs we extract only the objects for which we have at least three flares to obtain a good linear fit whilst including as many stars as possible. There are 31 objects in the SSO flaring sample (of 78 objects) which have at least three flares (the 31 objects have between 3--19 flares, with an average of 5 flares). We bin the energies per 0.1\,erg and calculate the cumulative frequency of flares greater than that energy. In Figure \ref{fig:ffd} we present the best-fit power laws from Equation \ref{eq:powerlaw}. We also present the individual FFDs and line fits for each of the 31 objects in Appendix \ref{ap:ffds}, to demonstrate the diversity of power law fits between stars of the same spectral type. We also performed the same FFD fitting using only the actual flare energies (unbinned) which did not change our results. We note that these results are not corrected for the detection efficiency as in Section \ref{senstivitymap} due to the complexity of applying it to the cumulative flare rates for low numbers of flares. Instead we correct for the recovery rate in the next section, where we combine flares from multiple stars to yield a much larger dataset.

By extrapolating the relationship between the flare energy and cumulative frequency for each star, we can predict the amount of energy that flares would deliver to the planets orbiting those stars. However, extrapolating the power-law relationship from our parameter space into the high-energy-flare regime is dangerously unreliable, as it is common to see breaks in these power-law relationships \citep{Silverberg20161243, Paudel2018Dwarfs}. If we instead treat these linear fits as upper limits, then we can give estimates for the maximum frequency of high-energy flares. We note that the power laws also vary with the energy binning and ranges chosen for fitting.

We see a tentative trend that as the stars get cooler, the flares we detect become less energetic and less frequent. However, we see a large diversity of FFD profiles even within the same spectral type. This could be due to variation in stellar age or metallicity. Since we are biased towards detecting lower energy flares for cooler stars, it is possible we more significantly underestimate the rates of lower energy flares for our earlier M stars. This effect may result in steeper power-law slopes, further separating the mid-M dwarfs from the late-M and L dwarfs. However, we only have one M4, M9 and L0 object with at least 3 flares, and within the larger sample of M5--M7 stars, we see little distinction. The decline in flare rates for cooler spectral types could be partially explained by our decreased sensitivity for early-M stars. If only the infrequent, high energy flares are within reach, we only pick up the objects that flare more often. Conversely, since we are more sensitive to detecting lower energy flares on late-M and L dwarfs, we are able to identify stars with lower flaring frequencies. This effect explains the lack of low frequency flaring stars of early spectral types, but not the lack of high frequency flaring stars of later spectral types. As we have a small sample of flaring ultra-cool stars (see Figure \ref{fig:flares_spec}), we cannot confirm whether they, as a whole, flare less frequently. \citet{Paudel2018Dwarfs} similarly find that the flare rates for low-energy flares decrease as they move towards later spectral types, with L0 and L1 dwarfs having the lowest flaring rates. Critically, however, they find shallower slopes for cool stars (also in agreement with \citet{Gizis2013KeplerFlares} and \citet{Mullan2018PhotosynthesisFlares}), implying that they have higher occurrence rates for the high-energy flares that are inaccessible to us. High frequency of these flares could be sufficient for a star to enter the abiogenesis zone. However, we are unable to confirm this trend, due to our small parameter space and the large uncertainties on our individual power law slopes.

If we take a more conservative threshold of at least 5 flares to fit the power law, then we extract FFDs for only M5-M7 stars (for which we have a much larger flare sample). For this sub-sample we find values for $\alpha$ in the range 1.2--2, roughly in agreement with \cite{Paudel2018Dwarfs}, who measure 1.3--2 with their sample of 10 UCDs (for a smaller sample, but a similar spectral type range). \citet{Paudel2018Dwarfs} discuss various reasons for the variations in FFD slopes that are not dependent on age or spectral type, such as rotation, stellar spot coverage, and magnetic field topology.

\subsubsection{Converting Bolometric Flare Energies to U-band}
We calculate the energy in the $U$ band, $E_{U}$, by integrating the flux density in the $U$-band spectral response function, as in \cite{Gunther2020Dwarfs}. Similarly to Section \ref{flareenergies}, we calculate:
\begin{equation} \label{eq:uband}
    E_{U} = \int_{\textrm{flare}}{A_{\textrm{flare}}~\textrm{d}t} \int_{U}{R_{\lambda} B_{\lambda}(T_{\textrm{flare}}) ~\textrm{d}\lambda}.
\end{equation}
Therefore,
\begin{equation} \label{eq:X_i+z_u}
    E_{U} = E\frac{1}{\sigma_{\textrm{SB}} T_{\textrm{flare}}^4}\int_{U}{R_{\lambda} B_{\lambda}(T_{\textrm{flare}}) ~\textrm{d}\lambda},
\end{equation}
where $R_{\lambda}$ is now the Johnson $U$-band response function. The normalisation of the response function is of greater importance here than when calculating the bolometric energies, as it does not cancel out. 
From this we estimate that 7.6 per cent of the flare's bolometric energy falls in the $U$ band. 

\subsubsection{Prebiotic chemistry}
Here we consider the laboratory work of \cite{Rimmer2018TheExoplanets} in defining ``abiogenesis zones'' around each of our potential planet hosts, outside of which it is unlikely a planet would receive enough energy for the following prebiotic photochemistry scenario to occur. Specifically, the pathway considered in this work is the synthesis of ribonucteotides, as a precursor to ribonucleic acid (RNA) \citep{Patel2015CommonProtometabolism,Sutherland2017Opinion:Beginning,Xu2018PhotochemicalFerrocyanide}. By considering stellar flares as a mechanism for providing this UV energy, we can calculate the abiogenesis zones around the stars in our sample from their FFDs. Whether or not this chemistry is possible on a planet does not tell us if life has originated there, but instead, whether this mechanism can allow that planet to generate this first building block for RNA. Conversely, if a planet does not receive the necessary energy for this reaction it does not rule out alternate prebiotic pathways for the origins of life. 

\cite{Gunther2020Dwarfs} determine the necessary flare frequency to power prebiotic chemistry for a planet receiving the same amount of flux from its host as the Earth, by adapting the abiogenesis zone equations from  \cite{Rimmer2018TheExoplanets} as follows:
\begin{equation} \label{eq:abiogenesis_zone}
    \nu \geq 25.5\textrm{day}^{-1}\left(\frac{10^{34}\textrm{\,erg}}{E_{\textrm{U}}}\right)\left(\frac{R_{*}}{R_{\sun}}\right)^2\left(\frac{T_{*}}{T_{\sun}}\right)^4.
\end{equation}
From Section \ref{ffd} we calculate the flare energies in the $U$-band from each flare's bolometric energy, using $E_{U} = 0.076 E$. 

If we extrapolate the FFD power laws to $E = 10^{34}$\,ergs, we have only one star that provides the necessary UV flux to reach the abiogensesis zone (see Figure \ref{fig:ffd}). This star is an M6 object. If we extrapolate the FFD power laws instead to $E = 10^{36}$\,ergs, then we have 13 objects that reach the abiogenesis zone (with spectral types consisting of 1 M4, 1M8, 8 M6 and 3 M7). However, as our sample is confined to probing the low energy flare regime, we have to be careful to not over-extrapolate to high energies. Therefore, we do not extend the power law predictions further than $E = 10^{34}$\,ergs. Almost all the FFDs will eventually reach the abiogenesis zone, but at very high energy the uncertainties in our power law fit would be too large to draw any conclusions.

\subsubsection{Photosynthesis}\label{photosynthesis}
\cite{Lingam2019PhotosynthesisStars} define a threshold for sustaining a biosphere on an Earth-like planet using flare-driven photosynthesis. By considering ``photosynthetically active radiation'' in the region of 400--750\,nm, they find a similar functional form to Equation \ref{eq:abiogenesis_zone} for the minimum flare rates necessary to receive the same photon flux on a temperate planet as on Earth:
\begin{equation} \label{eq:photosynthesis}
    \nu \geq 9.4\times10^{3}\textrm{day}^{-1}\left(\frac{10^{34}\textrm{\,erg}}{E}\right)\left(\frac{R_{*}}{R_{\sun}}\right)^2\left(\frac{T_{*}}{T_{\sun}}\right)^4.
\end{equation}
This threshold is a significantly greater inhibitor than that for prebiotic chemistry. We plot this threshold for each star with our FFDs in Figure \ref{fig:ffd}. None of our stars exceed this threshold when extrapolating to $E = 10^{34}$\,erg. Because the condition for oxygenic photosynthesis is more strict than Equation \ref{eq:abiogenesis_zone}, if this an object satisfies Equation \ref{eq:photosynthesis}, it must also meet the requirements for the abiogenesis zone.


\subsection{Applying our Sensitivity to the Flare Sample}\label{senstivitymap}

To compare the average flare occurrence rates for different spectral types, we must account for the incompleteness of our sample. Due to the difficulty of detecting small amplitude flares above photometric scatter, we likely underestimate the frequency of low energy flares. This is often seen as a non-linear `tail-off' from the expected power law in log-log space (Equation \ref{eq:powerlaw}). 

\citet{Jackman2021StellarSurvey} derive the following equation (equivalent to their Equation 3) for the number of flares, $N$, with energies, $E$, greater than $E_{\textrm{flare}}$: 
\begin{dmath} \label{eq:jackman_ngts}
    N(E \geq E_{\textrm{flare}})=\frac{k}{\alpha-1}\left( R(E_{\textrm{flare}})E_{\textrm{flare}}^{-\alpha+1} +   \int^{E_{\textrm{max}}}_{E_{\textrm{flare}}}{R'(E)E^{-\alpha+1} ~\textrm{d}E}\right),
\end{dmath}
where $k$ and $\alpha$ are the same constants in Equation \ref{eq:power_law2}, $R(E)$ is the flare recovery fraction, $R'(E)$ is the differentiated flare recovery fraction, and $E_{\textrm{max}}$ is the energy at which the recovery fraction saturates. For high energy flares, where $R=1$, Equation \ref{eq:jackman_ngts} reduces to the power law in Equation \ref{eq:power_law2}, whereas for low energy flares, where $R=0$, it will reduce to the constant value of the integral (equivalent to the `tail-off' effect).

Due to the small population of flares on the earliest and latest stars in our sample, we build up flare numbers by combining spectral types into the following five bins: M4--5, M6, M7, M8 and M9--L0. We calculate every star's unique recovery fraction, $R(E)$, based on the results of the injection-recovery tests. We bin flare energies into 12 logarithmically spaced bins from $10^{28}$ to $10^{34}$\,ergs. For every star we find a value for the recovery fraction in each energy bin by calculating the fraction of flares with energies in that bin that were recovered. We smooth the recovery fraction using a Wiener filter of three bins. We average the recovery fractions within each spectral type bin to get the recovery fraction of an ``average'' star. 

We then extract the observed flare occurrence rates from our flare sample. First, we isolate only the stars in each spectral type bin. We include all flaring and non-flaring stars in the SSO sample, to avoid overestimating our flaring rates. For every energy bin, $E$, we sum the total number of flares with energies $\geq E$. Then we divide the total number of flares by the sum of the stars' total observation times to produce flaring rates. In doing so we make the assumption that 10 stars observed for 10 hours is equivalent to 1 star observed for 100 hours. 

\begin{figure}
    \includegraphics[width=\columnwidth]{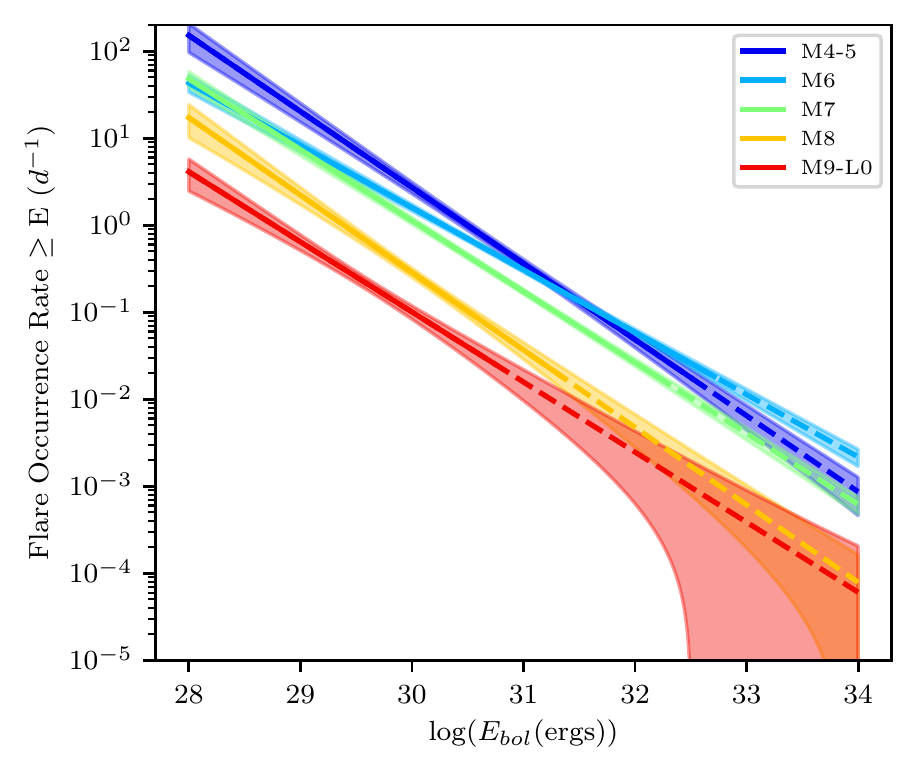}
     \caption{The computed flare occurrence rates (Equation \ref{eq:power_law2}), and their 1$\sigma$ uncertainties, against bolometric flare energy for an average star of each spectral type bin. These flare rates have been corrected for our incompleteness at low flare energies and represent the `intrinsic' flaring rate (equivalent to Equation \ref{eq:jackman_ngts} at very high energies). The solid lines mark up to the highest energy flare for each spectral type bin, after that the ``extrapolated'' region is marked by the dotted line.}
     \label{fig:MCMC}
\end{figure}

We fit Equation \ref{eq:jackman_ngts} to our observed flare occurrence rates and energies using a Markov Chain Monte Carlo (MCMC) parameter optimisation procedure. To implement MCMC we use the \textsc{emcee} Python package \citep{Foreman-Mackey2013EmceeHammer}. We used 32 walkers for 10000 steps and the last 2000 steps to sample the posterior distribution. As in \citet{Ilin2019Astrophysics67} and \citet{Jackman2021StellarSurvey}, we multiplied our errors by $R(E)^{-0.5}$ to account for larger uncertainties in recovering the smallest energy flares. The observed flare occurrence rates for each spectral type bin are presented in Table \ref{tab:flareoccurrences} and the results of the best-fit power laws for each spectral type bin are shown in Figure \ref{fig:MCMC}. We obtain similar values of $\alpha$ = $1.88\pm0.05$, $1.72\pm0.02$, $1.82\pm0.02$, $1.89\pm0.07$, and $1.81\pm0.08$ for M4--5, M6, M7, M8, and M9--L0 respectively. This work demonstrates that on average the power-law relationship does not change with spectral type in the mid- to late-M regime, despite the large variations within each spectral type. FFDs with similar gradients, but offset (different y-intercepts), implies that these spectral types produce similar relative proportions of high and low energy flares, but the coolest stars have lower rates of flares of all energies.

\setlength\extrarowheight{2pt}
\begin{table*}
 \centering
 \caption{The observed flare occurrence rates per day for each spectral type and flare energy bin. The parameter $\nu_{28}$ represents the rate of flares per day with $E\geq 10^{28}$\,ergs. There are no flares detected for any spectral type with $E\geq 10^{33}$\,ergs.}
 \label{tab:flareoccurrences}
 \begin{tabular}{|l|c|c|c|c|c|}
  \hline
  Spectral Type & $\nu_{28}$ & $\nu_{28.5}$ & $\nu_{29}$ & $\nu_{29.5}$ & $\nu_{30}$ \\
  \hline
  M4--M5 & $0.7\pm0.9$ & $0.7\pm0.6$ & $0.7\pm0.4$ & $0.7\pm0.2$ & $0.68\pm0.17$ \\
  M6 & $0.7\pm0.4$ & $0.7\pm0.2$ & $0.66\pm0.17$ & $0.64\pm0.11$ & $0.60\pm0.07$ \\
  M7 & $0.53\pm0.08$ & $0.6\pm0.5$ & $0.6\pm0.3$ & $0.58\pm0.16$ & $0.56\pm0.12$  \\
  M8 & $0.15\pm0.03$ & $0.2\pm0.2$ & $0.17\pm0.17$ & $0.17\pm0.07$ & $0.17\pm0.04$  \\
  M9--L0 & $0.052\pm0.009$ & $0.14\pm0.09$ & $0.14\pm0.05$ & $0.14\pm0.04$ & $0.10\pm0.02$\\ 
  \hline
  \\
  \hline
  Spectral Type & $\nu_{30.5}$ & $\nu_{31}$ & $\nu_{31.5}$ & $\nu_{32}$ & $\nu_{32.5}$ \\
  \hline
  M4--M5  & $0.44\pm0.08$ & $0.24\pm0.05$ & $0.055\pm0.014$ & $0.037\pm0.009$ & $0.018\pm0.007$ \\
  M6 & $0.43\pm0.05$ & $0.24\pm0.02$ & $0.112\pm0.009$ & $0.047\pm0.004$ & $0.024\pm0.003$ \\
  M7 & $0.30\pm0.04$ & $0.140\pm0.016$ & $0.060\pm0.008$ & $0.020\pm0.002$ & -- \\
  M8 & $0.068\pm0.014$ & $0.017\pm0.005$ & --  & -- & -- \\
  M9--L0 & $0.017\pm0.006$ & -- & --  & -- & -- \\
  \hline
 \end{tabular}
\end{table*}

\setlength\extrarowheight{2pt}
\begin{table*}
 \centering
 \caption{The best-fit power law $\alpha$, for each spectral type bin, as determined using MCMC.}
 \label{tab:flarealpha}
 \begin{tabular}{|l|c|}
  \hline
  Spectral Type & $\alpha$ \\
  \hline
  M4--M5 & $1.88\pm0.05$\\
  M6 & $1.72\pm0.02$\\
  M7 & $1.82\pm0.02$\\
  M8 & $1.89\pm0.07$\\
  M9--L0 & $1.81\pm0.08$\\
  \hline
 \end{tabular}
\end{table*}

Our results are consistent with three similar studies of the relation between flaring rate and energy for cool stars. As previously mentioned, \citet{Paudel2018Dwarfs} measured a range of $\alpha$ from 1.3--2.0 for 10 UCDs from K2 lightcurves. \citet{Raetz2020Rotation-activityData} compiled K2 lightcurves and derived a value for $\alpha$ of $1.83\pm0.05$ for FFDs of the coolest stars in their sample (with spectral types M5--M6). \citet{Gizis2017Superflare} found $\alpha$=1.8 for three UCDs. However, our results have noticeably lower values of $\alpha$ than several other works on mid-to-late M-dwarfs. From Kepler lightcurves, \citet{Yang2017TheField} and \citet{Yang2019TheMission} generated flare catalogues and found on average M dwarfs had a power law indices of $\alpha=2.09\pm0.10$ (for fully convective stars) and $\alpha=2.07\pm0.35$ respectively. For a sample of cool dwarfs observed by TESS, \citet{Medina2020FlarePc} found $\alpha=1.98\pm0.02$. \citet{Seli2021ActivityTESS} specifically isolated TESS's observations of stars similar in spectral type to TRAPPIST–1, for which they found $\alpha$=2.11. This work also modified the FFD for TRAPPIST-1 generated by \citet{Vida2017FrequentLife} to include recovery rate, updating their value of $\alpha$ from 1.59 to $2.03\pm0.02$. \citet{Lin2021EDEN:Curves} used EDEN and K2 lightcurves to analyse the flaring activity of the nearby active M dwarf Wolf 359, for which they found $\alpha=2.13\pm0.14$. We could find a lower value of $\alpha$ if the incompleteness of the sample is not fully described by Equation \ref{eq:jackman_ngts}. We see a slightly earlier tail-off at low energies compared to the \citet{Jackman2021StellarSurvey} model for all spectral types, which may be evidence of this. Alternatively, previous works may find higher values for $\alpha$ than this work as they have studied higher energy flares on UCD targets using space-based observations. We are limited in this study to only frequent, low energy flares due to our observing strategy. Therefore, it is possible that power laws may steepen at higher flaring energies.


This analysis agrees with the results of Section \ref{ffd}, which show that there is a decline in flaring rates as we move to the coolest stars. The injection-recovery tests demonstrate that if flares were present on the lowest-mass stars we would be able to detect them; therefore, these flares must occur too infrequently to be detected with our survey strategy.

\subsection{Flares and rotation}
From the SSO data sample, we recover 69 (24 A, 22 B, and 23 L) rotators, with periods ranging from 2.2 hours to 65 days. We also find 29 U and 60 N class objects. From here on we refer to `rotators' as only grade A, B and L. 

We find 41 targets for which we can detect both flares and rotation. Therefore, of the 78 flaring objects, described above, 53 per cent have clear rotation. Alternatively, we detect flares on 59 per cent of our rotators. It appears that while the fraction of rotators across all spectral types stays consistently between 20--50 per cent, there are increasing proportions of slow rotators for the later M9 and L0 stars.

\begin{figure}
    \includegraphics[width=\columnwidth]{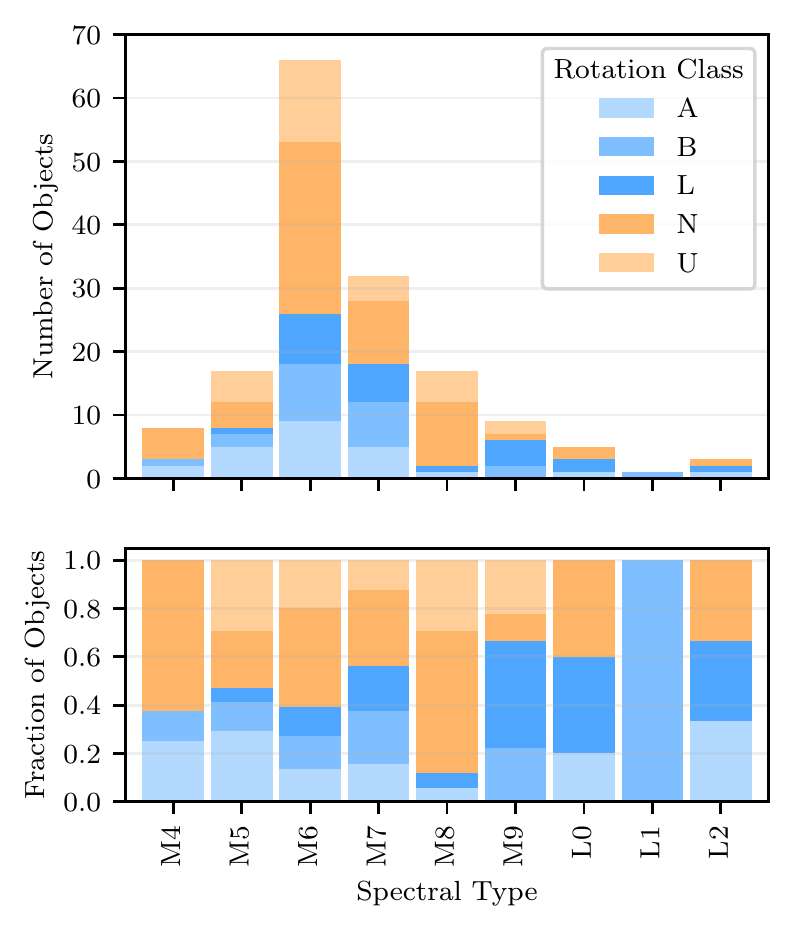}
     \caption{\textbf{Top:} Histogram of objects in the SSO data sample as a function of spectral type, divided by rotation class. The `rotators' are shown in varying shades of blue, and the `non-rotators' are shown in shades of orange. \textbf{Bottom:} Fraction of different rotator classes for each spectral type. }
     \label{fig:rot_spec}
\end{figure}

\begin{figure}
    \includegraphics[width=\columnwidth]{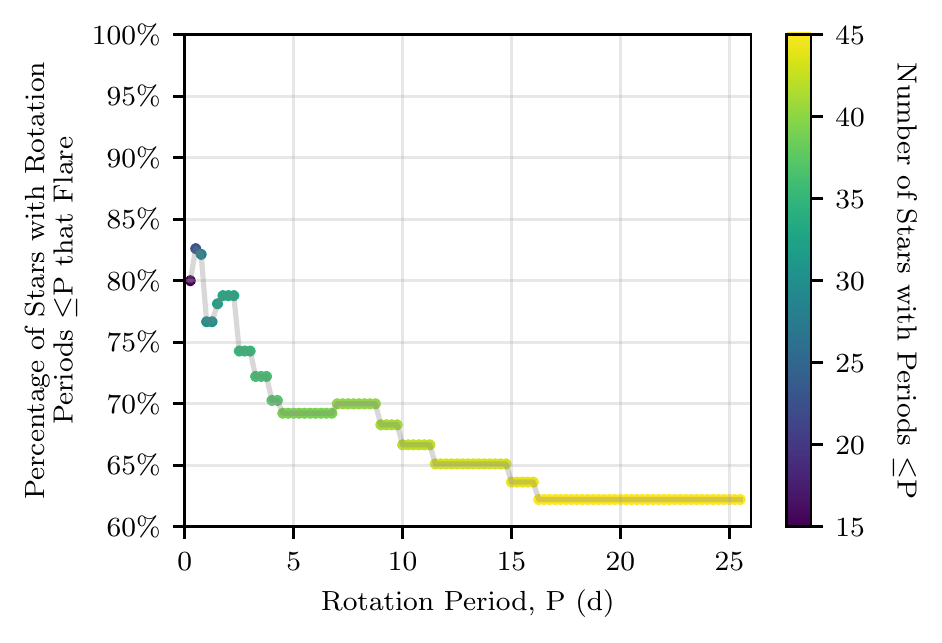}
     \caption{The proportion of stars with rotation periods $\leq P$ that have detected flares, against rotation period $P$, in steps of 0.25\,d. We only include our 46 `A' and `B' rotators, which range in period from 0.09\,d to 25.7\,d. The colour scale shows the number of stars with rotation period $\leq P$.}
     \label{fig:flares_rot}
\end{figure}

We see clearly that the very fast rotators are much more likely to flare than slow rotators. By comparing the rotation period in steps of 0.25\,d, $P$, with the proportion of stars with rotation periods $\leq P$ that flare (see Figure \ref{fig:flares_rot}), we demonstrate that the likelihood that we detect a star as flaring decreases as the rotation slows. In Figure \ref{fig:flares_rot} we only include our 46 `A' and `B' grade rotators, as the periods of the `L' (long period) grade rotators have very large uncertainties. In this sample at least 74 per cent of fast-rotating stars, with $P\leq2$\,d, flare. Comparatively, 59 per cent of all rotators flare, 63 per cent of `A' and `B' rotators flare, 42 per cent of stars with no detected rotation flare and 50 per cent of all stars flare. We note that the majority (33/46) of `A' and `B' rotators have periods $\leq2$\,d. 



\subsubsection{Comparison with MEarth}
\begin{figure}
    \includegraphics[width=\columnwidth]{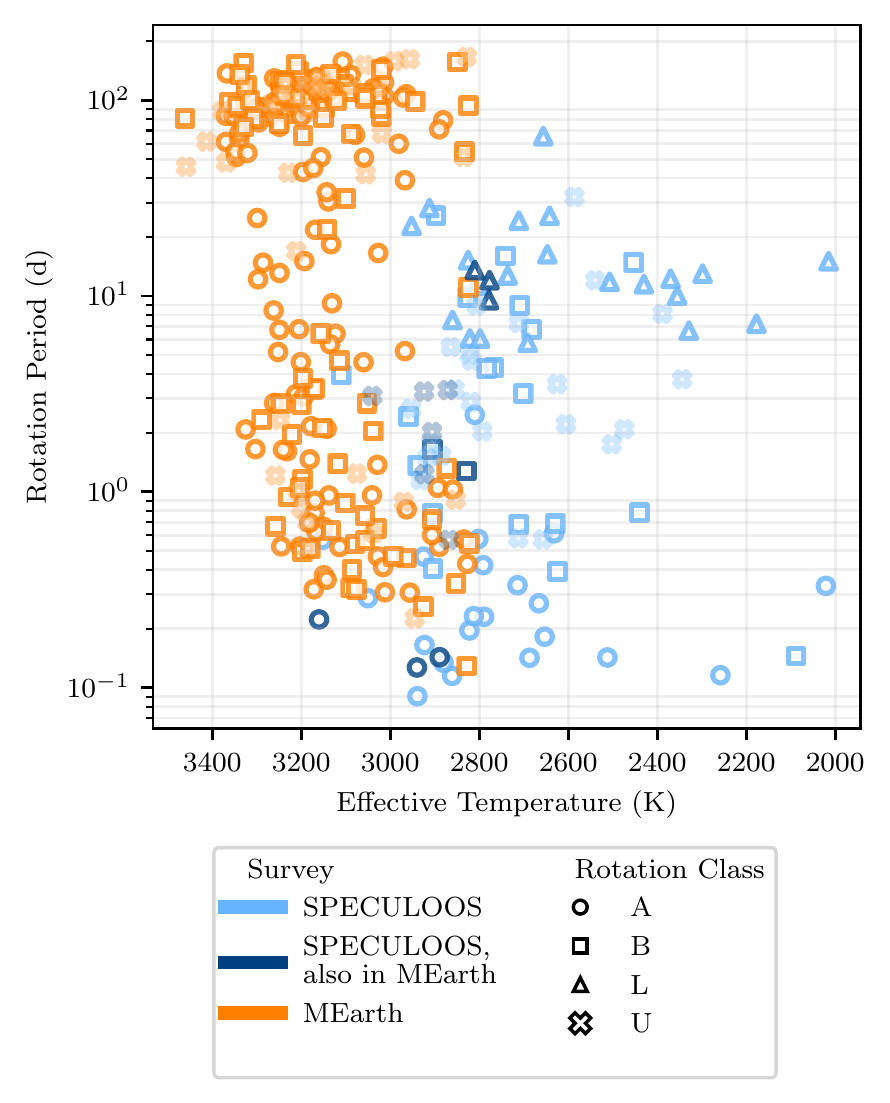}
     \caption{A comparison of the rotation periods measured in this SSO data sample with the rotation periods from MEarth \citep{Newton2018NewStars}. The objects observed only by SPECULOOS are in light blue and the objects observed only by MEarth are in light orange. The dark blue points are the measured period in SPECULOOS for objects that have also been observed by MEarth. The different rotation classes are indicated by the marker shapes, with the uncertain (U) rotation periods made transparent. Note that the L rotators, a class unique to SPECULOOS, provide only lower limit estimates of the periods, due to our shorter observing baselines.}
     \label{fig:mearth}
\end{figure}

Within the SSO Sample, 20 targets also appear in the \cite{Newton2018NewStars} MEarth-South rotation sample. Of these 20 objects, our results agree for eight, and they suggest periods not found by MEarth for another seven. For the eight objects in agreement: two have long period rotation measured by MEarth, with periods too long to be measured by SPECULOOS (for which we find no rotation); four have no detected rotation with either survey; and two have a similar measured period by both surveys (though one has only a tentative detection with the SSO). For the remaining 12 objects, five have clear, short periods detected by SPECULOOS but not by MEarth, and two have long-period, low-amplitude estimates given by SPECULOOS with no detection in MEarth. We discount the remaining five objects classified as U/N in both surveys, with missing or disagreeing periods. All of the rotation periods measured by the two surveys are shown in Figure \ref{fig:mearth}, and their overlapping observations are reported in Table \ref{tab:mearth}.

The difference in rotation periods measured by the two surveys is likely a result of their different observing strategies. While SPECULOOS performs continuous monitoring of every target for 4--8\,hours each night over several weeks, MEarth cycles through multiple targets during a night, returning to each at 20--30 minute cadences. This advantages SPECULOOS to observe very short, $<$5\,hour period rotators, and MEarth to measure very long, $>$50\,day rotation periods. We also derive a long-period rotation estimate with SPECULOOS for two objects with small amplitudes that may not be evident in the MEarth data, due to limitations on the lowest mass stars. MEarth experiences a drop-off in recovery rate for stars with $M<0.2M_{\sun}$ (which includes all of the objects in our sample) likely due to systematics and the precipitable water effect \citep{Newton2018NewStars}.

The \cite{Newton2018NewStars} sample shows a clear dichotomy of fast rotators, with periods less than 10 days, and slow rotators, with periods greater than 70 days. We are unable to confirm this, as the majority of our sample have been observed for less than a 10-day span. For the clear slow rotators, which we observe for less than a full phase, we can only assign a long-period estimate (L). The apparent gap in rotation periods from our sample between one and two days is likely a bias from the difficulty in disentangling real rotation from 1-day aliases.

\begin{table*}
 	\centering
 	\caption{Comparison of SPECULOOS and MEarth rotations for overlapping objects in rotation analyses of both surveys \citep{Newton2018NewStars}.}
 	\label{tab:mearth}
 	\rowcolors{2}{gray!10}{white}
 	\begin{tabular}{lcccc} 
 		\hline
 		Gaia ID & SSO Class & SSO Period & MEarth Class & MEarth Period\\
 		 & & (d) & & (d)\\
 		\hline
 		4127182375667696128 & U & 0.57 & A & 0.571\\
 		5072067381112863104 & L & 12.04 & B & 11.067\\
 		6056881391901174528 & N & N/A & B & 156.741\\
 		6421389047155380352 & N & N/A & U & 51.202\\
 		6340981796172195584 & A & 0.22 & N & N/A\\
 		6405457982659103872 & A & 0.14 & N & N/A\\
 		2631857350835259392 & A & 0.13 & N & N/A\\
 		5565156633450986752 & B & 1.65 & N & N/A\\
 		6914281796143286784 & B & 1.28 & N & N/A\\
 		6504700451938373760 & L & 13.56 & N & N/A\\
 		2393563872239260928 & L & 9.53 & N & N/A\\
 	    3005440443830195968 & U & 3.25 & U & 0.913\\
 		4349305645979265920 & U & 3.31 & N & N/A\\
 		4654435618927743872 & U & 1.23 & N & N/A\\
 		4854708878788267264 & U & 3.09 & N & N/A\\
 		6494861747014476288 & U & 2.02 & U & 166.165\\
 		6385548541499112448 & N & N/A & N & N/A\\
 		3175523485214138624 & N & N/A & N & N/A\\
 		3474993275382942208 & N & N/A & N & N/A\\
 		3562427951852172288 & N & N/A & N & N/A\\
 		\hline
 	\end{tabular}
 \end{table*}

\subsubsection{Comparison with TESS}
\citet{Seli2021ActivityTESS} study the relationships between age, activity and rotation for a sample of 248 ``TRAPPIST-1 analogues'' from 30-minute TESS Full Frame Image (FFI) observations. They detect a total of 94 flare events on 21 stars. We compare their target catalogue to the SSO data sample and find 35 objects that appear in both. Of these 35 objects, we detect 19 as flaring including five of the seven targets that \citet{Seli2021ActivityTESS} identify as flaring. The five objects that both surveys identify as flaring stars are Gaia DR2 2331849006126794880, 2349207644734247808, 3200303384927512960 (as shown in Figure \ref{fig:flare}), 4825880783419986432 and 5055805741577757824. The two flaring stars detected from TESS lightcurves, on which SPECULOOS does not detect flares are Gaia DR2 4967628688601251200 and 5637175400984142336. For 14 of the targets in this overlapping sample of 35 objects neither survey detects any flares. The 21 overlapping objects (excluding the 14 that have no flares detected by either study) are presented in Table \ref{tab:tess}.

For 42 stars in their sample \citet{Seli2021ActivityTESS} detect rotation periods. Out of the stars that appear in both UCD samples, \citet{Seli2021ActivityTESS} find rotation periods for four objects that appear in our catalogue. We find rotation periods for 16 of the 35 objects (with rotation classes A, B or L). Therefore, we can also compare both our flaring and rotation results for this handful of targets. For three of these objects we detect an alias of the period extracted from TESS 30-minute FFIs. For each of these three objects we inspected the SSO lightcurves and periodograms. In each case, the rotation period measured by \citet{Seli2021ActivityTESS} is the second most promising period, however, the peaks in the periodogram are very close in power, and the chosen period usually gives a better phase coverage. This is the reason for the `B' grade we assign to rotators Gaia DR2 2349207644734247808 and 4967628688601251200. Gaia DR2 3200303384927512960 is assigned a grade of `A', however, it is a unique case. For this target we detect a rotation period of 0.334\,d, while \citet{Seli2021ActivityTESS} measure a period of 0.50062\,d. Rotation periods of 0.33, 0.5 or 1\,d are special scenarios for ground-based observations where, over short baselines, we only see the same part of the phase on every observation night. Therefore, due to the limit imposed by the day-night cycle, we are unable to distinguish between these periods. The target for which we do not detect any rotation period, Gaia DR2 4971892010576979840, has an amplitude for periodic photometric variation of 2.1\,mmag (measured by \citet{Seli2021ActivityTESS}). Upon inspection of this object's SSO lightcurve this periodicity is too small to detect above the level of photometric scatter, and residual systematics.

Since TESS is a space-based mission which does not contend with the day-night cycle or ground-based systematics, and they perform 27\,day continuous monitoring, this provides a large number of phases for fast-rotating stars, which can counteract lower photometric precisions for red dwarfs. Therefore, we would expect rotation periods derived from TESS lightcurves to be more accurate than those from ground-based observations.


\begin{table*}
 	\centering
 	\caption{Comparison of this work and \citet{Seli2021ActivityTESS} for 21 overlapping objects with flares detected in either flaring analysis.}
 	\label{tab:tess}
 	\rowcolors{2}{gray!10}{white}
 	\begin{tabular}{lccccccc} 
 		\hline
 		Gaia ID & SSO Flaring? & SSO Period & SSO Amp. & SSO Rotation Class & TESS Flaring? & TESS Period & TESS Amp.\\
 		& & (d) & (mmag) & & & (d) & (mmag) \\
 		\hline
            2331849006126794880 & Y & N/A & N/A & N & Y & N/A & N/A\\
            2349207644734247808 & Y & 0.687 & 17.0 & B & Y & 2.13867 & 20.9\\
            3200303384927512960 & Y & 0.334 & 24.1 & A & Y & 0.50062 & 3.8\\
            4825880783419986432 & Y & 65.468 & 26.1 & L & Y & N/A & N/A\\
            5055805741577757824 & Y & 25.662 & 6.8 & L & Y & N/A & N/A\\
            4967628688601251200 & N & 0.682 & 6.5 & B & Y & 0.40391 & 4.2\\
            5637175400984142336 & N & 0.573 & 5.2 & U & Y & N/A & N/A\\
            
            3421840993510952192 & Y & 4.297 & 22.3 & B & N & N/A & N/A\\
            3493736924979792768 & Y & 0.142 & 8.8 & A & N & N/A & N/A\\
            3504014060164255104 & Y & 16.326 & 5.4 & L & N & N/A & N/A\\
            4404521333221783680 & Y & N/A & N/A & N & N & N/A & N/A\\
            4928644747924606848 & Y & 11.819 & 7.8 & L & N & N/A & N/A\\
            4971892010576979840 & Y & N/A & N/A & N & N & 0.70254 & 2.1\\
            5047423236725995136 & Y & N/A & N/A & N & N & N/A & N/A\\
            5156623295621846016 & Y & 6.752 & 3.9 & B & N & N/A & N/A\\
            5392287051645815168 & Y & 0.182 & 8.6 & A & N & N/A & N/A\\
            5469802724480366848 & Y & N/A & N/A & N & N & N/A & N/A\\
            56252256123908096 & Y & 0.614 & 29.4 & A & N & N/A & N/A\\
            5657734928392398976 & Y & N/A & N/A & N & N & N/A & N/A\\
            6357834388848708224 & Y & 2.034 & 15.7 & U & N & N/A & N/A\\
            6733860940302404864 & Y & 0.27 & 4.4 & A & N & N/A & N/A\\
            
 		\hline
 	\end{tabular}
 \end{table*}

For their catalogue of stars, \citet{Seli2021ActivityTESS} find that $\sim$8 per cent are flaring. This flaring fraction contrasts this work, where we detect flares on 28 of the 57 M7--M9 objects within the SSO sample (49 per cent). There may be several reasons for this discrepancy. Firstly, \citet{Seli2021ActivityTESS} detect flares from lightcurves with a low time resolution of 30 minutes. TESS is also not optimised for very red objects such as UCDs, therefore the photometric precision is lower than for SPECULOOS. Both of these factors inhibit a flare survey on TESS data from detecting low energy flares (which have small amplitudes and short durations) on UCDs. Instead, continuous 27-day monitoring of each TESS field allows \citet{Seli2021ActivityTESS} to probe the rarer, high energy (even superflare) flare regime. This is demonstrated by their recovery fraction, which is less than 20 per cent for flares with energies below $10^{31.5}$\,ergs, compared to SSO recovery fractions of 50--70 per cent (depending on spectral type) for the same energy. Additionally, in this paper we demonstrate that cooler stars flare less frequently at all energies, therefore it is likely that a survey focused on high energy flares on cool stars would detect significantly fewer flares. Finally, space-based data is less impacted than ground-based data by time-varying systematics due to the Earth's rapidly-changing atmosphere, therefore the risk of false positives is presumably lower.


\section{Discussion}\label{discussion}
\subsection{Extension into the ultra-cool regime}

The breadth of the SSO flare sample allows us to compare the flaring and rotating behaviour of mid-M stars with ultra-cool dwarfs. We predominantly detect flares on M5--M7 stars ($\sim$60-70 per cent flare), and decreasing proportions of flaring stars for both earlier and later M dwarfs. Since there is a spectral dependency to our flare detection (Figure \ref{fig:recover_energies_teff}), it is likely that we are missing the lower energy flares on mid-M dwarfs, and therefore underestimating the proportion of flaring stars. On the other hand, these higher mass objects do produce rarer, high-energy flares more frequently than late-M dwarfs \citep{Lacy1976UVData.}. These high-energy flares should be easier to detect. Our results also agree with other surveys, with around $\sim30$ per cent of M4 stars flaring. It is possible that these surveys are limited in different ways for mid-M stars, which exist at the redder end of their samples, rather than at the bluer end of ours.

\citet{Lacy1976UVData.} showed that the flares on mid-M dwarfs were more frequent and lower energy than on early-M dwarfs. \cite{Hilton2010MSpectra} found that this trend continued into the late-M dwarf regime (M6--M8). From our findings we see little difference between spectral types up to M7. However, our results extend beyond M8, where we find that flares of all energies appear to become less frequent. From our injection-recovery tests, we conclude that if higher energy flares were present on the reddest stars in our sample, then we would detect them; therefore, the lack of high-energy flares for late-M and L dwarfs on the sampled timescales of 0.8 to 8.7 days is likely real. 

We see a relationship between flaring activity and rotation, especially for our very fastest rotators. The stars with rapid rotation periods of $<$5\,days, are much more likely to flare than the rest of the SSO sample. At least 74 per cent of the stars with rotation periods faster than 2 days flare, compared to around 59 per cent of all rotating stars and 42 per cent of non-rotating stars. Therefore, we are more likely to detect flares on stars where we can determine a rotation period than stars we cannot. Possible explanations are that fast rotators flare much more frequently, more energetically, or are more likely to flare than stars with no detectable rotation period. Being unable to detect a rotation period may be a symptom that a star has a period too long to be measured by our survey, or the photometric variations on the rotation timescale are too small to be seen, possibly due to small spot covering fractions, a low spot-photosphere contrast, or a particular orientation of spots (e.g. distributed at the poles). This contrast between rotators and non-rotators is further enhanced for the very fastest periods ($P\leq2$\,d). Assessing the spectral distribution of our fastest rotators shows that almost all of the M4--M7 stars with periods less than 2 days flare. Comparatively, very few objects with spectral types later than M8 flare, fast-rotating or otherwise. Stars with fast rotation will have more magnetic energy available; therefore, these results are consistent with faster rotators exhibiting more active behaviours \citep{Davenport2016THEFLARES,Allred2015AFlares, Newton2017ROTATION,Yang2019TheMission, Gunther2020Dwarfs, Martinez2019AASAS-SN}. Specific activity-rotation studies on cool stars, such as those carried out by \citet{West2015AnDwarfs}, \citet{Stelzer2016AData}, \citet{Medina2020FlarePc}, and \citet{Raetz2020Rotation-activityData}, have also found that fast rotating stars flare more frequently than slow rotators. \citet{Seli2021ActivityTESS} found that $\sim$50 per cent of their fast-rotating ($P<5$\,days) late-M stars flared, compared to $\sim$70 per cent of our sample with the same periods.

In this work, we have compared results for SPECULOOS targets that also appear in the rotation study performed by \citet{Newton2018NewStars} on MEarth data, and in the activity-rotation study from \citet{Seli2021ActivityTESS} on TESS observations. SPECULOOS and MEarth are two ground-based surveys focused on cool stars, but with very different observing strategies. SPECULOOS is a fast-cadence survey that targets ultra-cool dwarfs for relatively short baselines of 4--8\,days. MEarth, however, revisits its M dwarf targets only every 20--30\,minutes, over a period of several months. With SPECULOOS's lightcurves, therefore, we are able to study the short-term photometric variability of an object, such as its flaring and fast rotation (<10\,d), whereas with MEarth's lightcurves we can monitor an object's long-term variability, such as any slow rotation (>50\,d) and stellar surface evolution. SPECULOOS and MEarth, however, are both ground-based surveys which have to contest with the challenges of observing through the Earth's atmosphere and the loss of data from the day-night cycle. TESS, a space telescope, is able to perform continuous 27\,d monitoring of each field, however ultra-cool dwarfs are not its ideal targets. Therefore, TESS lightcurves of UCDs are mostly only available from the 30-minute FFIs, and have generally lower precisions. SPECULOOS has the short exposure times and high precisions necessary to identify frequent low energy flares, however for rotation, the irregular data sampling and large gaps (limited by weather conditions and the observing cycle) make it difficult to identify the correct period above its aliases. For this reason we recommend caution when interpreting the rotation results from this survey, and provide `rotation classes' to help identify targets with rotation period uncertainties. TESS is only able to detect the rarer high energy flares (with large amplitudes and slow decay times), however, its ability to provide continuous phase coverage of fast-rotating objects, and to observe up to hundreds of cycles, make it well suited for rotation studies. Comparisons such as these, between ground and space-based surveys, and between fast and slow-cadence observing strategies highlight that they are highly complementary. Leveraging the differences and the overlaps between different surveys allows us to fully explore the photometric activity parameter space of UCDs, and unearth more information about the subsequent impact on any potential planets they might host.

\subsection{Planetary Impact}
Often used to put stellar flares into context, the largest solar flare ever directly observed on Earth was the `Carrington Event' in 1859 \citep{Carrington1859Description1859, Hodgson1859OnSun}. This flare released an energy of $10^{32}$\,erg and the associated coronal mass ejection hit the Earth's magnetosphere, resulting in the largest geomagnetic storm on record. This storm disrupted telegraph systems and produced auroras reaching almost as far as the Equator. While our flare sample is relatively low-energy compared to the wealth of recorded flares on other stellar types, in our sample we detect 14 flares with energies greater than $10^{32}$\,erg on 10 different stars. Characterising these interactions between planets and their host stars, especially those which have the capability to initiate or destroy life, is vital to our understanding of planetary habitability.

\subsubsection{Habitability of Earth-sized planets around UCDs}

From calculating FFDs for the stars in our sample, we can place energetic bounds on whether flares can drive photosynthesis (Section \ref{photosynthesis}) or provide their planets with enough UV flux for the prebiotic chemistry described in Section \ref{ffd} to occur. If we extrapolate the FFD power laws to $E=10^{34}$\,erg, we have only one M6 star that provides this necessary UV flux. However, there are several caveats to these results. As we see a non-linear ``tail-off'' effect for low energy flares due to our detection limits we likely underestimate low energy flares rates, resulting in shallower FFD power laws in Section \ref{ffd}. Therefore, without accounting for the completeness of our flare sample this work only places upper bounds on the stars whose planets can reach the abiogensesis zone, or which can sustain photosynthesis. As we move towards cooler stars, we see that the frequency of flaring decreases (see Figure \ref{fig:ffd}). However, there are only one of each of the M4, M9 and L0 stars, with at least 3 flares detected. Due to our low detection efficiency for mid-M dwarfs, we are biased to detect only the most frequently flaring stars; therefore, it is likely that we would see an artificial increase in flare rate for M4 objects. In contrast, we should be more sensitive to flares of all energies on our late-M and L dwarf stars, so this bias does not explain the decrease in flare frequency for these stars. 

By including our detection sensitivity in Section \ref{senstivitymap} and considering an ``average'' star for each spectral type, we can conclude more general relationships between flare frequency and spectral type. While there is a wide diversity of flare frequencies for individual stars' within a given spectral type, on average cooler stars have lower flaring rates. If this trend holds for a larger sample of UCDs, it may have severe consequences on the search for life around the very lowest mass objects. Similar conclusions have been reached by work on the flaring activity of TRAPPIST-1 \citep{Glazier2020EvryscopeHabitability,Ducrot2020TRAPPIST-1:Worlds, Seli2021ActivityTESS} and ``TRAPPIST-1 analogues'' \citep{Seli2021ActivityTESS}.

If the majority of late-M and L dwarfs have flaring rates too low to initiate the synthesis of ribonucleotides, then for life to originate it must happen through a different mechanism. Possible alternatives are surface hydrothermal vents \citep{Rimmer2019OriginVents}, aerial biospheres \citep{Lingam2019BrownLife}, impact-shock synthesis \citep{Furukawa2015NucleobaseOcean} or extraterrestrial delivery \citep{Rimmer2019OriginVents, Krijt2017FastSystem}. We also find that none of our sample meets the requirement set out by \cite{Lingam2019PhotosynthesisStars} to sustain an Earth-like biosphere, due to the significantly lower photosynthetically active radiation (with wavelengths of 400--750\,nm) incident within UCD habitable zones. These results are consistent with the work of \cite{Mullan2018PhotosynthesisFlares} and \cite{Lingam2019PhotosynthesisStars}. However, we are considering an Earth-centric viewpoint, and it has been suggested that photosynthesis could proceed using deep-sea hydrothermal vents \citep{Beatty2005AnVent} or at infrared wavelengths, where UCDs are brightest \citep{Scalo2007MDetection, Takizawa2017Red-edgeM-dwarfs, Claudi2021Super-EarthsLab}.

The activity of M dwarfs is predicted to decrease as they age \citep{West2008CONSTRAININGSAMPLE, West2015AnDwarfs, Paudel2019K2Dwarfs}. Therefore, it is possible that earlier in these stars' life cycles they would produce more energetic and frequent flares, with enough UV flux to trigger prebiotic chemistry on their planets, or even for photosynthesis. The results in this paper only represent a snapshot of each object at its current point in time. We do not consider age in this work due to the huge uncertainties on stellar age estimates for these objects.

Modelling of the impact of stellar flares on modern Earth-like analogues by \cite{Tilley2019ModelingPlanet} shows that stars with at least 0.1 flare per day with energies above $10^{34}$\,erg will strip the ozone layers from any terrestrial planets they host. We do not extend our FFDs to energies above $10^{34}$\,erg as we are probing the low energy flare regime; therefore, we do not consider ozone depletion in our analysis.

Here we have only discussed two aspects of how flares could impact the habitability of planets around UCDs. In reality, we would need to apply a more holistic approach. There are a multitude of other factors (such as orbit dynamics, stellar variability, tidal locking, atmospheric composition, stellar age, etc.) that influence the environments on planets hosted by UCDs, and other possible pathways for life that haven't been considered here.



\section{Conclusions}\label{conclusions}

By analysing the high-precision photometry produced by SPECULOOS, we can monitor the activity of late-M and ultra-cool dwarfs. This allows us to probe further into the ultra-cool regime, expanding our understanding of the relationships between activity and rotation into a new parameter space. This work can be used in conjunction with previous and ongoing flare studies on earlier M dwarfs, such as those carried out on TESS, K2, and MEarth observations, to assess the diversity of M dwarfs and how well we can extrapolate the results of these studies to lower mass objects. We demonstrate the benefit of a large-scale flare study of UCDs, such as this, to extending existing flare catalogues.

Using our flare sample to make more general predictions about flaring activity as we move to very low mass stars, we address two different scenarios in which flares may assist the origin and sustenance of life on planets they host. Firstly, the quiescent UV flux of cool stars is too low for the synthesis of ribonucleotides, a major step in prebiotic chemistry. Secondly, the visible light incident on planets around cool stars is also found to be too low for photosynthesis to occur. In both cases, while we find it is possible that flares on active M dwarfs could provide the necessary extra radiation, the frequency of flares of all energies appears to decrease with spectral type. These two considerations alone suggest UCD systems may not be favourable sites for abiogenesis, though other factors---known and unknown---certainly impact the likelihood of finding life in these systems as well.

\section*{Acknowledgements}
The research leading to these results has received funding from the European Research Council (ERC) under the FP/2007--2013 ERC grant agreement n$^{\circ}$ 336480, and under the European Union's Horizon 2020 research and innovation programme (grants agreements n$^{\circ}$ 679030 \& 803193/BEBOP); from an Actions de Recherche Concert\'{e}e (ARC) grant, financed by the Wallonia--Brussels Federation, from the Balzan Prize Foundation, from the BELSPO/BRAIN2.0 research program (PORTAL project), from the Science and Technology Facilities Council (STFC; grant n$^\circ$ ST/S00193X/1), and from F.R.S-FNRS (Research Project ID T010920F). This work was also partially supported by a grant from the Simons Foundation (PI Queloz, grant number 327127), as well as by the MERAC foundation (PI Triaud). PI Gillon is F.R.S.-FNRS Senior Research Associate.
MNG acknowledges support from MIT's Kavli Institute as a Juan Carlos Torres Fellow.
B.-O. D. acknowledges support from the Swiss National Science Foundation (PP00P2-163967 and PP00P2-190080). This research is also supported by funding from STFC grants n${^\circ}$ ST/S00193X/1 and ST/S00305/1.
Finally, we would like to thank the anonymous referee for their helpful, detailed comments.

\section*{Data Availability}
The data underlying this article is not currently public. The SPECULOOS-South Consortium will make all SPECULOOS-South Facility reduced data products available to the ESO Science Archive Facility following the regular Phase 3 process as described at \url{http://www.eso.org/sci/observing/phase3.html}.

\twocolumn

\bibliographystyle{mnras}
\bibliography{references2} 

\begin{thebibliography}{}
\makeatletter
\relax
\def\mn@urlcharsother{\let\do\@makeother \do\$\do\&\do\#\do\^\do\_\do\%\do\~}
\def\mn@doi{\begingroup\mn@urlcharsother \@ifnextchar [ {\mn@doi@}
  {\mn@doi@[]}}
\def\mn@doi@[#1]#2{\def\@tempa{#1}\ifx\@tempa\@empty \href
  {http://dx.doi.org/#2} {doi:#2}\else \href {http://dx.doi.org/#2} {#1}\fi
  \endgroup}
\def\mn@eprint#1#2{\mn@eprint@#1:#2::\@nil}
\def\mn@eprint@arXiv#1{\href {http://arxiv.org/abs/#1} {{\tt arXiv:#1}}}
\def\mn@eprint@dblp#1{\href {http://dblp.uni-trier.de/rec/bibtex/#1.xml}
  {dblp:#1}}
\def\mn@eprint@#1:#2:#3:#4\@nil{\def\@tempa {#1}\def\@tempb {#2}\def\@tempc
  {#3}\ifx \@tempc \@empty \let \@tempc \@tempb \let \@tempb \@tempa \fi \ifx
  \@tempb \@empty \def\@tempb {arXiv}\fi \@ifundefined
  {mn@eprint@\@tempb}{\@tempb:\@tempc}{\expandafter \expandafter \csname
  mn@eprint@\@tempb\endcsname \expandafter{\@tempc}}}

\bibitem[\protect\citeauthoryear{Allred, Kowalski  \& Carlsson}{Allred
  et~al.}{2015}]{Allred2015AFlares}
Allred J.~C.,  Kowalski A.~F.,   Carlsson M.,  2015, \mn@doi [Astrophysical
  Journal] {10.1088/0004-637X/809/1/104}, 809

\bibitem[\protect\citeauthoryear{Anfinogentov, Nakariakov, Mathioudakis,
  Van~Doorsselaere  \& Kowalski}{Anfinogentov
  et~al.}{2013}]{Anfinogentov2013THEMEGAFLARE}
Anfinogentov S.,  Nakariakov V.~M.,  Mathioudakis M.,  Van~Doorsselaere T.,
  Kowalski A.~F.,  2013, \mn@doi [The Astrophysical Journal]
  {10.1088/0004-637X/773/2/156}, 773, 156

\bibitem[\protect\citeauthoryear{Ballard \& Johnson}{Ballard \&
  Johnson}{2016}]{Ballard2016PLANETS}
Ballard S.,  Johnson J.~A.,  2016, \mn@doi [The Astrophysical Journal]
  {10.3847/0004-637x/816/2/66}, 816, 66

\bibitem[\protect\citeauthoryear{Beatty et~al.,}{Beatty
  et~al.}{2005}]{Beatty2005AnVent}
Beatty J.~T.,  et~al., 2005, \mn@doi [Proceedings of the National Academy of
  Science] {10.1073/pnas.0503674102}, 102, 9306

\bibitem[\protect\citeauthoryear{Benz \& G{\"{u}}del}{Benz \&
  G{\"{u}}del}{2010}]{Benz2010PhysicalObjects}
Benz A.~O.,  G{\"{u}}del M.,  2010, \mn@doi [Annual Review of Astronomy and
  Astrophysics] {10.1146/annurev-astro-082708-101757}, 48, 241

\bibitem[\protect\citeauthoryear{Borucki et~al.,}{Borucki
  et~al.}{2010}]{Borucki2010KeplerResults}
Borucki W.~J.,  et~al., 2010, \mn@doi [Science] {10.1126/science.1185402}, 327,
  977

\bibitem[\protect\citeauthoryear{Broeg, Fern{\'{a}}ndez  \&
  Neuh{\aa}user}{Broeg et~al.}{2005}]{Broeg2005AStar}
Broeg C.,  Fern{\'{a}}ndez M.,   Neuh{\aa}user R.,  2005, \mn@doi
  [Astronomische Nachrichten] {10.1002/asna.200410350}, 326, 134

\bibitem[\protect\citeauthoryear{Buccino, Lemarchand  \& Mauas}{Buccino
  et~al.}{2007}]{Buccino2007UVStars}
Buccino A.~P.,  Lemarchand G.~A.,   Mauas P.~J.,  2007, \mn@doi [Icarus]
  {10.1016/j.icarus.2007.08.012}, 192, 582

\bibitem[\protect\citeauthoryear{Burdanov, Delrez, Gillon  \& Jehin}{Burdanov
  et~al.}{2018}]{Burdanov2018SPECULOOSTRAPPIST}
Burdanov A.,  Delrez L.,  Gillon M.,   Jehin E.,  2018, in Deeg H.~J.,
  Belmonte J.~A.,  eds, , Handbook of Exoplanets.
p.~130, \mn@doi{10.1007/978-3-319-55333-7{\_}130}

\bibitem[\protect\citeauthoryear{Cantrell, Henry  \& White}{Cantrell
  et~al.}{2013}]{Cantrell2013TheNeighbors}
Cantrell J.~R.,  Henry T.~J.,   White R.~J.,  2013, \mn@doi [The Astronomical
  Journal] {10.1088/0004-6256/146/4/99}, 146, 99

\bibitem[\protect\citeauthoryear{Carrington}{Carrington}{1859}]{Carrington1859Description1859}
Carrington R.~C.,  1859, \mn@doi [Monthly Notices of the Royal Astronomical
  Society] {10.1093/mnras/20.1.13}, 20, 13

\bibitem[\protect\citeauthoryear{Chabrier}{Chabrier}{2003}]{Chabrier2003GalacticFunction}
Chabrier G.,  2003, \mn@doi [Publications of the Astronomical Society of the
  Pacific] {10.1086/376392}, 115, 763

\bibitem[\protect\citeauthoryear{Chabrier \& Baraffe}{Chabrier \&
  Baraffe}{1997}]{Chabrier1997StructureStars}
Chabrier G.,  Baraffe I.,  1997, Astronomy and Astrophysics, 327, 1039

\bibitem[\protect\citeauthoryear{Claudi et~al.,}{Claudi
  et~al.}{2021}]{Claudi2021Super-EarthsLab}
Claudi R.,  et~al., 2021, \mn@doi [Life] {10.3390/life11010010}, 11, 1

\bibitem[\protect\citeauthoryear{Covone, Ienco, Cacciapuoti  \& Inno}{Covone
  et~al.}{2021}]{Covone2021EfficiencyZone}
Covone G.,  Ienco R.~M.,  Cacciapuoti L.,   Inno L.,  2021, \mn@doi [Monthly
  Notices of the Royal Astronomical Society] {10.1093/mnras/stab1357}, 505,
  3329

\bibitem[\protect\citeauthoryear{Davenport}{Davenport}{2016}]{Davenport2016THEFLARES}
Davenport J. R.~A.,  2016, \mn@doi [The Astrophysical Journal]
  {10.3847/0004-637X/829/1/23}, 829

\bibitem[\protect\citeauthoryear{Davenport et~al.,}{Davenport
  et~al.}{2014}]{Davenport2014Kepler1243}
Davenport J.~R.,  et~al., 2014, \mn@doi [Astrophysical Journal]
  {10.1088/0004-637X/797/2/122}, 797, 122

\bibitem[\protect\citeauthoryear{Davenport, Covey, Clarke, Boeck, Cornet  \&
  Hawley}{Davenport et~al.}{2019}]{Davenport2019TheAge}
Davenport J. R.~A.,  Covey K.~R.,  Clarke R.~W.,  Boeck A.~C.,  Cornet J.,
  Hawley S.~L.,  2019, \mn@doi [The Astrophysical Journal]
  {10.3847/1538-4357/aafb76}, 871, 241

\bibitem[\protect\citeauthoryear{Delrez et~al.,}{Delrez
  et~al.}{2018}]{Delrez2018SPECULOOS:Dwarfs}
Delrez L.,  et~al., 2018, Ground-based and Airborne Telescopes VII, 10700, 49

\bibitem[\protect\citeauthoryear{Demory et~al.,}{Demory
  et~al.}{2020}]{Demory2020ATOI-1266}
Demory B.~O.,  et~al., 2020, \mn@doi [Astronomy and Astrophysics]
  {10.1051/0004-6361/202038616}, 642, A49

\bibitem[\protect\citeauthoryear{Dressing \& Charbonneau}{Dressing \&
  Charbonneau}{2015}]{Dressing2015THESENSITIVITY}
Dressing C.~D.,  Charbonneau D.,  2015, \mn@doi [The Astrophysical Journal]
  {10.1088/0004-637X/807/1/45}, 807, 45

\bibitem[\protect\citeauthoryear{Ducrot et~al.,}{Ducrot
  et~al.}{2020}]{Ducrot2020TRAPPIST-1:Worlds}
Ducrot E.,  et~al., 2020, \mn@doi [Astronomy and Astrophysics]
  {10.1051/0004-6361/201937392}, 640

\bibitem[\protect\citeauthoryear{Foreman-Mackey, Hogg, Lang  \&
  Goodman}{Foreman-Mackey et~al.}{2013}]{Foreman-Mackey2013EmceeHammer}
Foreman-Mackey D.,  Hogg D.~W.,  Lang D.,   Goodman J.,  2013, \mn@doi
  [Publications of the Astronomical Society of the Pacific] {10.1086/670067},
  125, 306

\bibitem[\protect\citeauthoryear{Furukawa, Nakazawa, Sekine, Kobayashi  \&
  Kakegawa}{Furukawa et~al.}{2015}]{Furukawa2015NucleobaseOcean}
Furukawa Y.,  Nakazawa H.,  Sekine T.,  Kobayashi T.,   Kakegawa T.,  2015,
  \mn@doi [Earth and Planetary Science Letters] {10.1016/j.epsl.2015.07.049},
  429, 216

\bibitem[\protect\citeauthoryear{Gardner et~al.,}{Gardner
  et~al.}{2006}]{Gardner2006TheTelescope}
Gardner J.~P.,  et~al., 2006, \mn@doi [Space Science Reviews]
  {10.1007/s11214-006-8315-7}, 123, 485

\bibitem[\protect\citeauthoryear{Gershberg}{Gershberg}{1972}]{Gershberg1972Some1967-71}
Gershberg R.~E.,  1972, \mn@doi [Astrophysics and Space Science]
  {10.1007/BF00643168}, 19, 75

\bibitem[\protect\citeauthoryear{Gillon}{Gillon}{2018}]{Gillon2018SearchingWorlds}
Gillon M.,  2018, \mn@doi [Nature Astronomy] {10.1038/s41550-018-0443-y}, 2,
  344

\bibitem[\protect\citeauthoryear{Gillon, Jehin, Magain, Chantry,
  Hutsem{\'{e}}kers, Manfroid, Queloz  \& Udry}{Gillon
  et~al.}{2011}]{Gillon2011TRAPPIST:Systems}
Gillon M.,  Jehin E.,  Magain P.,  Chantry V.,  Hutsem{\'{e}}kers D.,  Manfroid
  J.,  Queloz D.,   Udry S.,  2011, \mn@doi [EPJ Web of Conferences]
  {10.1051/epjconf/20101106002}, 11, 06002

\bibitem[\protect\citeauthoryear{Gizis, Burgasser, Berger, Williams, Vrba, Cruz
   \& Metchev}{Gizis et~al.}{2013}]{Gizis2013KeplerFlares}
Gizis J.~E.,  Burgasser A.~J.,  Berger E.,  Williams P. K.~G.,  Vrba F.~J.,
  Cruz K.~L.,   Metchev S.,  2013, \mn@doi [Astrophysical Journal]
  {10.1088/0004-637X/779/2/172}, 779

\bibitem[\protect\citeauthoryear{Gizis, Paudel, Schmidt, Williams  \&
  Burgasser}{Gizis et~al.}{2017}]{Gizis2017Superflare}
Gizis J.~E.,  Paudel R.~R.,  Schmidt S.~J.,  Williams P. K.~G.,   Burgasser
  A.~J.,  2017, \mn@doi [The Astrophysical Journal] {10.3847/1538-4357/aa6197},
  838, 22

\bibitem[\protect\citeauthoryear{Glazier, Howard, Corbett, Law, Ratzloff, Fors
  \& Ser}{Glazier et~al.}{2020}]{Glazier2020EvryscopeHabitability}
Glazier A.~L.,  Howard W.~S.,  Corbett H.,  Law N.~M.,  Ratzloff J.~K.,  Fors
  O.,   Ser D.~d.,  2020, \mn@doi [The Astrophysical Journal]
  {10.3847/1538-4357/ABA4A6}, 900, 27

\bibitem[\protect\citeauthoryear{G{\"{u}}nther et~al.,}{G{\"{u}}nther
  et~al.}{2020}]{Gunther2020Dwarfs}
G{\"{u}}nther M.~N.,  et~al., 2020, \mn@doi [The Astronomical Journal]
  {10.3847/1538-3881/ab5d3a}, 159, 60

\bibitem[\protect\citeauthoryear{Hawley, Davenport, Kowalski, Wisniewski, Hebb,
  Deitrick  \& Hilton}{Hawley et~al.}{2014}]{Hawley2014KeplerDwarfs}
Hawley S.~L.,  Davenport J.~R.,  Kowalski A.~F.,  Wisniewski J.~P.,  Hebb L.,
  Deitrick R.,   Hilton E.~J.,  2014, \mn@doi [Astrophysical Journal]
  {10.1088/0004-637X/797/2/121}, 797, 121

\bibitem[\protect\citeauthoryear{Henry}{Henry}{2004}]{Henry2004TheEnd}
Henry T.~J.,  2004, in Spectroscopically and Spatially Resolving the Components
  of the Close Binary Stars. pp 159--165

\bibitem[\protect\citeauthoryear{Hilton, West, Hawley  \& Kowalski}{Hilton
  et~al.}{2010}]{Hilton2010MSpectra}
Hilton E.~J.,  West A.~A.,  Hawley S.~L.,   Kowalski A.~F.,  2010, \mn@doi [The
  Astronomical Journal] {10.1088/0004-6256/140/5/1402}, 140, 1402

\bibitem[\protect\citeauthoryear{Hodgson}{Hodgson}{1859}]{Hodgson1859OnSun}
Hodgson R.,  1859, \mn@doi [Monthly Notices of the Royal Astronomical Society]
  {10.1093/mnras/20.1.15a}, 20, 15

\bibitem[\protect\citeauthoryear{Howard, Corbett, Law, Ratzloff, Glazier, Fors,
  Ser  \& Haislip}{Howard et~al.}{2019}]{Howard2019EvryFlare.Sky}
Howard W.~S.,  Corbett H.,  Law N.~M.,  Ratzloff J.~K.,  Glazier A.,  Fors O.,
  Ser D.~d.,   Haislip J.,  2019, \mn@doi [The Astrophysical Journal]
  {10.3847/1538-4357/ab2767}, 881, 9

\bibitem[\protect\citeauthoryear{Howard et~al.,}{Howard
  et~al.}{2020a}]{Howard2020EvryFlare.Sky}
Howard W.~S.,  et~al., 2020a, \mn@doi [The Astrophysical Journal]
  {10.3847/1538-4357/ab9081}, 895, 140

\bibitem[\protect\citeauthoryear{Howard et~al.,}{Howard
  et~al.}{2020b}]{Howard2020EvryFlare.TESS}
Howard W.~S.,  et~al., 2020b, \mn@doi [The Astrophysical Journal]
  {10.3847/1538-4357/abb5b4}, 902, 115

\bibitem[\protect\citeauthoryear{Howell et~al.,}{Howell
  et~al.}{2014}]{Howell2014TheResults}
Howell S.~B.,  et~al., 2014, \mn@doi [Publications of the Astronomical Society
  of the Pacific] {10.1086/676406}, 126, 398

\bibitem[\protect\citeauthoryear{Ilin, Schmidt, Davenport  \& Strassmeier}{Ilin
  et~al.}{2019}]{Ilin2019Astrophysics67}
Ilin E.,  Schmidt S.~J.,  Davenport J. R.~A.,   Strassmeier K.~G.,  2019,
  \mn@doi [Astronomy {\&} Astrophysics] {10.1051/0004-6361/201834400}, 622, 133

\bibitem[\protect\citeauthoryear{Jackman et~al.,}{Jackman
  et~al.}{2019}]{Jackman2019DetectionSurvey}
Jackman J.~A.,  et~al., 2019, \mn@doi [Monthly Notices of the Royal
  Astronomical Society: Letters] {10.1093/mnrasl/slz039}, 485, L136

\bibitem[\protect\citeauthoryear{Jackman et~al.,}{Jackman
  et~al.}{2021}]{Jackman2021StellarSurvey}
Jackman J. A.~G.,  et~al., 2021, \mn@doi [Monthly Notices of the Royal
  Astronomical Society] {10.1093/mnras/stab979}, 000, 1

\bibitem[\protect\citeauthoryear{Jehin et~al.,}{Jehin
  et~al.}{2011}]{Jehin2011TRAPPIST:Telescope}
Jehin E.,  et~al., 2011, The Messenger, 145, 2

\bibitem[\protect\citeauthoryear{Jehin et~al.,}{Jehin
  et~al.}{2018}]{Jehin2018ThePlanets}
Jehin E.,  et~al., 2018, \mn@doi [The Messenger] {10.18727/0722-6691/5105},
  174, 2

\bibitem[\protect\citeauthoryear{Kaltenegger \& Traub}{Kaltenegger \&
  Traub}{2009}]{Kaltenegger2009TransitsPlanets}
Kaltenegger L.,  Traub W.~A.,  2009, \mn@doi [Astrophysical Journal]
  {10.1088/0004-637X/698/1/519}, 698, 519

\bibitem[\protect\citeauthoryear{Kerber et~al.,}{Kerber
  et~al.}{2012}]{Kerber2012AObservatory}
Kerber F.,  et~al., 2012, in McLean I.~S.,  Ramsay S.~K.,   Takami H.,  eds,
  Society of Photo-Optical Instrumentation Engineers (SPIE) Conference Series
  Vol. 8446, Ground-based and Airborne Instrumentation for Astronomy IV. p.
  84463N, \mn@doi{10.1117/12.924340}

\bibitem[\protect\citeauthoryear{Kowalski, Hawley, Wisniewski, Osten, Hilton,
  Holtzman, Schmidt  \& Davenport}{Kowalski
  et~al.}{2013}]{Kowalski2013Time-resolvedSpectra}
Kowalski A.~F.,  Hawley S.~L.,  Wisniewski J.~P.,  Osten R.~A.,  Hilton E.~J.,
  Holtzman J.~A.,  Schmidt S.~J.,   Davenport J.~R.,  2013, \mn@doi
  [Astrophysical Journal, Supplement Series] {10.1088/0067-0049/207/1/15}, 207,
  15

\bibitem[\protect\citeauthoryear{Krijt, Bowling, Lyons  \& Ciesla}{Krijt
  et~al.}{2017}]{Krijt2017FastSystem}
Krijt S.,  Bowling T.~J.,  Lyons R.~J.,   Ciesla F.~J.,  2017, \mn@doi [The
  Astrophysical Journal] {10.3847/2041-8213/aa6b9f}, 839, L21

\bibitem[\protect\citeauthoryear{Lacy, Moffett  \& Evans}{Lacy
  et~al.}{1976}]{Lacy1976UVData.}
Lacy C.~H.,  Moffett T.~J.,   Evans D.~S.,  1976, \mn@doi [Astrophysical
  Journal Supplement Series] {10.1086/190358}, 30, 85

\bibitem[\protect\citeauthoryear{Lammer et~al.,}{Lammer
  et~al.}{2007}]{Lammer2007CoronalZones}
Lammer H.,  et~al., 2007, \mn@doi [Astrobiology] {10.1089/ast.2006.0128}, 7,
  185

\bibitem[\protect\citeauthoryear{Law et~al.,}{Law
  et~al.}{2015}]{Law2015EvryscopeTelescopes}
Law N.~M.,  et~al., 2015, \mn@doi [Publications of the Astronomical Society of
  the Pacific] {10.1086/680521}, 127, 234

\bibitem[\protect\citeauthoryear{Lehmer, Catling, Parenteau  \& Hoehler}{Lehmer
  et~al.}{2018}]{Lehmer2018TheStars}
Lehmer O.~R.,  Catling D.~C.,  Parenteau M.~N.,   Hoehler T.~M.,  2018, \mn@doi
  [The Astrophysical Journal] {10.3847/1538-4357/aac104}, 859, 171

\bibitem[\protect\citeauthoryear{L{\'{e}}pine \& Bongiorno}{L{\'{e}}pine \&
  Bongiorno}{2007}]{Lepine2007Systems}
L{\'{e}}pine S.,  Bongiorno B.,  2007, \mn@doi [The Astronomical Journal]
  {10.1086/510333/FULLTEXT/}, 133, 889

\bibitem[\protect\citeauthoryear{Lienhard et~al.,}{Lienhard
  et~al.}{2020}]{Lienhard2020GlobalSurvey}
Lienhard F.,  et~al., 2020, \mn@doi [Monthly Notices of the Royal Astronomical
  Society] {10.1093/mnras/staa2054}, 497, 3790

\bibitem[\protect\citeauthoryear{Lin et~al.,}{Lin
  et~al.}{2021}]{Lin2021EDEN:Curves}
Lin C.-L.,  et~al., 2021, \mn@doi [The Astronomical Journal]
  {10.3847/1538-3881/ABF933}, 162, 11

\bibitem[\protect\citeauthoryear{Lingam \& Loeb}{Lingam \&
  Loeb}{2019a}]{Lingam2019PhotosynthesisStars}
Lingam M.,  Loeb A.,  2019a, \mn@doi [Monthly Notices of the Royal Astronomical
  Society] {10.1093/mnras/stz847}, 485, 5924

\bibitem[\protect\citeauthoryear{Lingam \& Loeb}{Lingam \&
  Loeb}{2019b}]{Lingam2019BrownLife}
Lingam M.,  Loeb A.,  2019b, \mn@doi [The Astrophysical Journal]
  {10.3847/1538-4357/ab3f35}, 883, 143

\bibitem[\protect\citeauthoryear{Lomb}{Lomb}{1976}]{Lomb1976Least-SquaresData}
Lomb N.~R.,  1976, \mn@doi [Astrophysics and Space Science]
  {10.1007/BF00648343}, 39, 447

\bibitem[\protect\citeauthoryear{Lurie, Davenport, Hawley, Wilkinson,
  Wisniewski, Kowalski  \& Hebb}{Lurie et~al.}{2015}]{Lurie2015KeplerB}
Lurie J.~C.,  Davenport J.~R.,  Hawley S.~L.,  Wilkinson T.~D.,  Wisniewski
  J.~P.,  Kowalski A.~F.,   Hebb L.,  2015, \mn@doi [Astrophysical Journal]
  {10.1088/0004-637X/800/2/95}, 800, 95

\bibitem[\protect\citeauthoryear{Mart{\'{i}}nez, Lopez, Shappee, Schmidt,
  Jayasinghe, Kochanek, Auchettl  \& Holoien}{Mart{\'{i}}nez
  et~al.}{2019}]{Martinez2019AASAS-SN}
Mart{\'{i}}nez R.~R.,  Lopez L.~A.,  Shappee B.~J.,  Schmidt S.~J.,  Jayasinghe
  T.,  Kochanek C.~S.,  Auchettl K.,   Holoien T. W.~S.,  2019, \mn@doi [The
  Astrophysical Journal] {10.3847/1538-4357/ab793a}, 892, 144

\bibitem[\protect\citeauthoryear{McCormac, Pollacco, Skillen, Faedi, Todd  \&
  Watson}{McCormac et~al.}{2013}]{McCormac2013DONUTS:Stars}
McCormac J.,  Pollacco D.,  Skillen I.,  Faedi F.,  Todd I.,   Watson C.~A.,
  2013, \mn@doi [Publications of the Astronomical Society of the Pacific]
  {10.1086/670940}, 125, 548

\bibitem[\protect\citeauthoryear{Medina, Winters, Irwin  \& Charbonneau}{Medina
  et~al.}{2020}]{Medina2020FlarePc}
Medina A.~A.,  Winters J.~G.,  Irwin J.~M.,   Charbonneau D.,  2020, \mn@doi
  [The Astrophysical Journal] {10.3847/1538-4357/abc686}, 905, 107

\bibitem[\protect\citeauthoryear{Moffett}{Moffett}{1974}]{Moffett1974UVDATA}
Moffett T.~J.,  1974, Technical Report~273, {UV CETI FLARE STARS: OBSERVATIONAL
  DATA}

\bibitem[\protect\citeauthoryear{Mondrik, Newton, Charbonneau  \&
  Irwin}{Mondrik et~al.}{2018}]{Mondrik2018AnDwarfs}
Mondrik N.,  Newton E.,  Charbonneau D.,   Irwin J.,  2018, \mn@doi [The
  Astrophysical Journal] {10.3847/1538-4357/aaee64}, 870, 10

\bibitem[\protect\citeauthoryear{Mullan \& Bais}{Mullan \&
  Bais}{2018}]{Mullan2018PhotosynthesisFlares}
Mullan D.~J.,  Bais H.~P.,  2018, \mn@doi [The Astrophysical Journal]
  {10.3847/1538-4357/aadfd1}, 865, 101

\bibitem[\protect\citeauthoryear{Murray et~al.,}{Murray
  et~al.}{2020}]{Murray2020PhotometrySPECULOOS-South}
Murray C.~A.,  et~al., 2020, \mn@doi [Monthly Notices of the Royal Astronomical
  Society] {10.1093/mnras/staa1283}, 495, 2446

\bibitem[\protect\citeauthoryear{Namekata et~al.,}{Namekata
  et~al.}{2017}]{Namekata2017StatisticalStars}
Namekata K.,  et~al., 2017, \mn@doi [The Astrophysical Journal]
  {10.3847/1538-4357/AA9B34}, 851, 91

\bibitem[\protect\citeauthoryear{Newton, Irwin, Charbonneau, Berta-Thomspon  \&
  West}{Newton et~al.}{2016}]{Newton2016RotationProject}
Newton E.~R.,  Irwin J.,  Charbonneau D.,  Berta-Thomspon Z.~K.,   West A.~A.,
  2016, in Proceedings of the International Astronomical Union. Cambridge
  University Press, pp 124--125, \mn@doi{10.1017/S174392131500602X}

\bibitem[\protect\citeauthoryear{Newton, Irwin, Charbonneau, Berlind, Calkins
  \& Mink}{Newton et~al.}{2017}]{Newton2017ROTATION}
Newton E.~R.,  Irwin J.,  Charbonneau D.,  Berlind P.,  Calkins M.~L.,   Mink
  J.,  2017, \mn@doi [The Astrophysical Journal] {10.3847/1538-4357/834/1/85},
  834, 85

\bibitem[\protect\citeauthoryear{Newton, Mondrik, Irwin, Winters  \&
  Charbonneau}{Newton et~al.}{2018}]{Newton2018NewStars}
Newton E.~R.,  Mondrik N.,  Irwin J.,  Winters J.~G.,   Charbonneau D.,  2018,
  \mn@doi [The Astronomical Journal] {10.3847/1538-3881/aad73b}, 156, 217

\bibitem[\protect\citeauthoryear{Niraula et~al.,}{Niraula
  et~al.}{2020}]{Niraula2020Team}
Niraula P.,  et~al., 2020, \mn@doi [The Astronomical Journal]
  {10.3847/1538-3881/aba95f}, 160, 172

\bibitem[\protect\citeauthoryear{Noyes, Hartmann, Baliunas, Duncan  \&
  Vaughan}{Noyes et~al.}{1984}]{Noyes1984RotationStars.}
Noyes R.~W.,  Hartmann L.~W.,  Baliunas S.~L.,  Duncan D.~K.,   Vaughan A.~H.,
  1984, \mn@doi [Astrophysical Journal] {10.1086/161945}, 279, 763

\bibitem[\protect\citeauthoryear{Nutzman \& Charbonneau}{Nutzman \&
  Charbonneau}{2008}]{Nutzman2008DesignDwarfs}
Nutzman P.,  Charbonneau D.,  2008, \mn@doi [Publications of the Astronomical
  Society of the Pacific] {10.1086/533420}, 120, 317

\bibitem[\protect\citeauthoryear{Patel, Percivalle, Ritson, Duffy  \&
  Sutherland}{Patel et~al.}{2015}]{Patel2015CommonProtometabolism}
Patel B.~H.,  Percivalle C.,  Ritson D.~J.,  Duffy C.~D.,   Sutherland J.~D.,
  2015, \mn@doi [Nature Chemistry] {10.1038/nchem.2202}, 7, 301

\bibitem[\protect\citeauthoryear{Paudel, Gizis, Mullan, Schmidt, Burgasser,
  Williams  \& Berger}{Paudel et~al.}{2018}]{Paudel2018Dwarfs}
Paudel R.~R.,  Gizis J.~E.,  Mullan D.~J.,  Schmidt S.~J.,  Burgasser A.~J.,
  Williams P. K.~G.,   Berger E.,  2018, \mn@doi [The Astrophysical Journal]
  {10.3847/1538-4357/aab8fe}, 858, 55

\bibitem[\protect\citeauthoryear{Paudel, Gizis, Mullan, Schmidt, Burgasser,
  Williams, Youngblood  \& Stassun}{Paudel et~al.}{2019}]{Paudel2019K2Dwarfs}
Paudel R.~R.,  Gizis J.~E.,  Mullan D.~J.,  Schmidt S.~J.,  Burgasser A.~J.,
  Williams P.~K.,  Youngblood A.,   Stassun K.~G.,  2019, \mn@doi [Monthly
  Notices of the Royal Astronomical Society] {10.1093/mnras/stz886}, 486, 1438

\bibitem[\protect\citeauthoryear{Raetz, Stelzer, Damasso  \& Scholz}{Raetz
  et~al.}{2020}]{Raetz2020Rotation-activityData}
Raetz S.,  Stelzer B.,  Damasso M.,   Scholz A.,  2020, \mn@doi [Astronomy {\&}
  Astrophysics] {10.1051/0004-6361/201937350}, 637, A22

\bibitem[\protect\citeauthoryear{Ranjan, Wordsworth  \& Sasselov}{Ranjan
  et~al.}{2017}]{Ranjan2017TheFollow-up}
Ranjan S.,  Wordsworth R.,   Sasselov D.~D.,  2017, \mn@doi [The Astrophysical
  Journal] {10.3847/1538-4357/aa773e}, 843, 110

\bibitem[\protect\citeauthoryear{Ricker et~al.,}{Ricker
  et~al.}{2015}]{Ricker2015TransitingTESS}
Ricker G.~R.,  et~al., 2015, \mn@doi [Journal of Astronomical Telescopes,
  Instruments, and Systems] {10.1117/1.JATIS.1.1.014003}, 1, 14003

\bibitem[\protect\citeauthoryear{Rimmer \& Shorttle}{Rimmer \&
  Shorttle}{2019}]{Rimmer2019OriginVents}
Rimmer P.~B.,  Shorttle O.,  2019, \mn@doi [Life] {10.3390/life9010012}, 9, 12

\bibitem[\protect\citeauthoryear{Rimmer, Xu, Thompson, Gillen, Sutherland  \&
  Queloz}{Rimmer et~al.}{2018}]{Rimmer2018TheExoplanets}
Rimmer P.~B.,  Xu J.,  Thompson S.~J.,  Gillen E.,  Sutherland J.~D.,   Queloz
  D.,  2018, \mn@doi [Science Advances] {10.1126/sciadv.aar3302}, 4, eaar3302

\bibitem[\protect\citeauthoryear{Scalo et~al.,}{Scalo
  et~al.}{2007}]{Scalo2007MDetection}
Scalo J.,  et~al., 2007, \mn@doi [Astrobiology] {10.1089/ast.2006.0125}, 7, 85

\bibitem[\protect\citeauthoryear{Scargle}{Scargle}{1982}]{Scargle1982StudiesData}
Scargle J.~D.,  1982, \mn@doi [The Astrophysical Journal] {10.1086/160554},
  263, 835

\bibitem[\protect\citeauthoryear{Schmidt et~al.,}{Schmidt
  et~al.}{2019}]{Schmidt2019TheASAS-SN}
Schmidt S.~J.,  et~al., 2019, \mn@doi [The Astrophysical Journal]
  {10.3847/1538-4357/ab148d}, 876, 115

\bibitem[\protect\citeauthoryear{Seager, Deming  \& Valenti}{Seager
  et~al.}{2009}]{Seager2009TransitingJWST}
Seager S.,  Deming D.,   Valenti J.~A.,  2009, \mn@doi [Astrophysics and Space
  Science Proceedings] {10.1007/978-1-4020-9457-6{\_}5}, 10, 123

\bibitem[\protect\citeauthoryear{Sebastian et~al.,}{Sebastian
  et~al.}{2021}]{Sebastian2021SPECULOOS:Strategy}
Sebastian D.,  et~al., 2021, \mn@doi [Astronomy {\&} Astrophysics]
  {10.1051/0004-6361/202038827}, 645, A100

\bibitem[\protect\citeauthoryear{Segura, Walkowicz, Meadows, Kasting  \&
  Hawley}{Segura et~al.}{2010}]{Segura2010TheDwarf}
Segura A.,  Walkowicz L.~M.,  Meadows V.,  Kasting J.,   Hawley S.,  2010,
  \mn@doi [Astrobiology] {10.1089/ast.2009.0376}, 10, 751

\bibitem[\protect\citeauthoryear{Seli, Vida, Mo{\'{o}}r, P{\'{a}}l  \&
  Ol{\'{a}}h}{Seli et~al.}{2021}]{Seli2021ActivityTESS}
Seli B.,  Vida K.,  Mo{\'{o}}r A.,  P{\'{a}}l A.,   Ol{\'{a}}h K.,  2021,
  \mn@doi [Astronomy {\&} Astrophysics] {10.1051/0004-6361/202040098}, 650,
  A138

\bibitem[\protect\citeauthoryear{Shappee et~al.,}{Shappee
  et~al.}{2014}]{Shappee2014THE2617}
Shappee B.~J.,  et~al., 2014, \mn@doi [Astrophysical Journal]
  {10.1088/0004-637X/788/1/48}, 788, 48

\bibitem[\protect\citeauthoryear{Shibayama et~al.,}{Shibayama
  et~al.}{2013}]{Shibayama2013SuperflaresSuperflares}
Shibayama T.,  et~al., 2013, \mn@doi [Astrophysical Journal, Supplement Series]
  {10.1088/0067-0049/209/1/5}, 209, 5

\bibitem[\protect\citeauthoryear{Silverberg, Kowalski, Davenport, Wisniewski,
  Hawley  \& Hilton}{Silverberg et~al.}{2016}]{Silverberg20161243}
Silverberg S.~M.,  Kowalski A.~F.,  Davenport J. R.~A.,  Wisniewski J.~P.,
  Hawley S.~L.,   Hilton E.~J.,  2016, \mn@doi [The Astrophysical Journal]
  {10.3847/0004-637x/829/2/129}, 829, 129

\bibitem[\protect\citeauthoryear{Skumanich}{Skumanich}{1972}]{Skumanich1972TIMEDEPLETION}
Skumanich A.,  1972, \mn@doi [The Astrophysical Journal] {10.1086/151310}, 171,
  565

\bibitem[\protect\citeauthoryear{Skumanich}{Skumanich}{1986}]{Skumanich1986SOMESTARS}
Skumanich A.,  1986, \mn@doi [The Astrophysical Journal] {10.1086/164654}, 309,
  858

\bibitem[\protect\citeauthoryear{Stelzer, Damasso, Scholz  \& Matt}{Stelzer
  et~al.}{2016}]{Stelzer2016AData}
Stelzer B.,  Damasso M.,  Scholz A.,   Matt S.~P.,  2016, \mn@doi [Monthly
  Notices of the Royal Astronomical Society] {10.1093/MNRAS/STW1936}, 463, 1844

\bibitem[\protect\citeauthoryear{Sutherland}{Sutherland}{2017}]{Sutherland2017Opinion:Beginning}
Sutherland J.~D.,  2017, \mn@doi [Nature Reviews Chemistry]
  {10.1038/s41570-016-0012}, 1, 0012

\bibitem[\protect\citeauthoryear{Takizawa, Minagawa, Tamura, Kusakabe  \&
  Narita}{Takizawa et~al.}{2017}]{Takizawa2017Red-edgeM-dwarfs}
Takizawa K.,  Minagawa J.,  Tamura M.,  Kusakabe N.,   Narita N.,  2017,
  \mn@doi [Scientific Reports] {10.1038/s41598-017-07948-5}, 7, 1

\bibitem[\protect\citeauthoryear{Tilley, Segura, Meadows, Hawley  \&
  Davenport}{Tilley et~al.}{2019}]{Tilley2019ModelingPlanet}
Tilley M.~A.,  Segura A.,  Meadows V.,  Hawley S.,   Davenport J.,  2019,
  \mn@doi [Astrobiology] {10.1089/ast.2017.1794}, 19, 64

\bibitem[\protect\citeauthoryear{Vida, K{\H{o}}v{\'{a}}ri, P{\'{a}}l,
  Ol{\'{a}}h  \& Kriskovics}{Vida et~al.}{2017}]{Vida2017FrequentLife}
Vida K.,  K{\H{o}}v{\'{a}}ri Z.,  P{\'{a}}l A.,  Ol{\'{a}}h K.,   Kriskovics
  L.,  2017, \mn@doi [The Astrophysical Journal] {10.3847/1538-4357/AA6F05},
  841, 124

\bibitem[\protect\citeauthoryear{West et~al.,}{West
  et~al.}{2004}]{West2004SPECTROSCOPICSUBDWARFS}
West A.~A.,  et~al., 2004, \mn@doi [The Astronomical Journal] {10.1086/421364},
  128, 426

\bibitem[\protect\citeauthoryear{West, Hawley, Bochanski, Covey, Reid, Dhital,
  Hilton  \& Masuda}{West et~al.}{2008}]{West2008CONSTRAININGSAMPLE}
West A.~A.,  Hawley S.~L.,  Bochanski J.~J.,  Covey K.~R.,  Reid I.~N.,  Dhital
  S.,  Hilton E.~J.,   Masuda M.,  2008, \mn@doi [The Astronomical Journal]
  {10.1088/0004-6256/135/3/785}, 135, 785

\bibitem[\protect\citeauthoryear{West, Weisenburger, Irwin, Berta-Thompson,
  Charbonneau, Dittmann  \& Pineda}{West et~al.}{2015}]{West2015AnDwarfs}
West A.~A.,  Weisenburger K.~L.,  Irwin J.,  Berta-Thompson Z.~K.,  Charbonneau
  D.,  Dittmann J.,   Pineda J.~S.,  2015, \mn@doi [Astrophysical Journal]
  {10.1088/0004-637X/812/1/3}, 812

\bibitem[\protect\citeauthoryear{Wheatley et~al.,}{Wheatley
  et~al.}{2018}]{Wheatley2018TheNGTS}
Wheatley P.~J.,  et~al., 2018, \mn@doi [Monthly Notices of the Royal
  Astronomical Society] {10.1093/mnras/stx2836}, 475, 4476

\bibitem[\protect\citeauthoryear{Williams, Berger, Irwin, Berta-Thompson  \&
  Charbonneau}{Williams
  et~al.}{2015}]{Williams2015SimultaneousJ13142039+1320011}
Williams P.~K.,  Berger E.,  Irwin J.,  Berta-Thompson Z.~K.,   Charbonneau D.,
   2015, \mn@doi [Astrophysical Journal] {10.1088/0004-637X/799/2/192}, 799,
  192

\bibitem[\protect\citeauthoryear{Wright \& Drake}{Wright \&
  Drake}{2016}]{Wright2016Solar-typeTachocline}
Wright N.~J.,  Drake J.~J.,  2016, \mn@doi [Nature] {10.1038/nature18638}, 535,
  526

\bibitem[\protect\citeauthoryear{Wright, Newton, Williams, Drake  \&
  Yadav}{Wright et~al.}{2018}]{Wright2018TheDwarfs}
Wright N.~J.,  Newton E.~R.,  Williams P.~K.,  Drake J.~J.,   Yadav R.~K.,
  2018, \mn@doi [Monthly Notices of the Royal Astronomical Society]
  {10.1093/MNRAS/STY1670}, 479, 2351

\bibitem[\protect\citeauthoryear{Xu, Ritson, Ranjan, Todd, Sasselov  \&
  Sutherland}{Xu et~al.}{2018}]{Xu2018PhotochemicalFerrocyanide}
Xu J.,  Ritson D.~J.,  Ranjan S.,  Todd Z.~R.,  Sasselov D.~D.,   Sutherland
  J.~D.,  2018, \mn@doi [Chemical Communications] {10.1039/c8cc01499j}, 54,
  5566

\bibitem[\protect\citeauthoryear{Yang \& Liu}{Yang \&
  Liu}{2019}]{Yang2019TheMission}
Yang H.,  Liu J.,  2019, \mn@doi [The Astrophysical Journal Supplement Series]
  {10.3847/1538-4365/AB0D28}, 241, 29

\bibitem[\protect\citeauthoryear{Yang et~al.,}{Yang
  et~al.}{2017}]{Yang2017TheField}
Yang H.,  et~al., 2017, \mn@doi [The Astrophysical Journal]
  {10.3847/1538-4357/AA8EA2}, 849, 36

\bibitem[\protect\citeauthoryear{de Wit \& Seager}{de~Wit \&
  Seager}{2013}]{deWit2013ConstrainingSpectroscopy}
de Wit J.,  Seager S.,  2013, \mn@doi [Science] {10.1126/science.1245450}, 342,
  1473

\makeatother
\end{thebibliography}




\appendix

\section{Flare Frequency Distributions}\label{ap:ffds}
In this appendix, we present the FFDs for each object with at least three flares. This includes the binned cumulative frequency against bolometric flare energy data and the best fit lines.
\begin{figure*}
	\includegraphics[width=\textwidth]{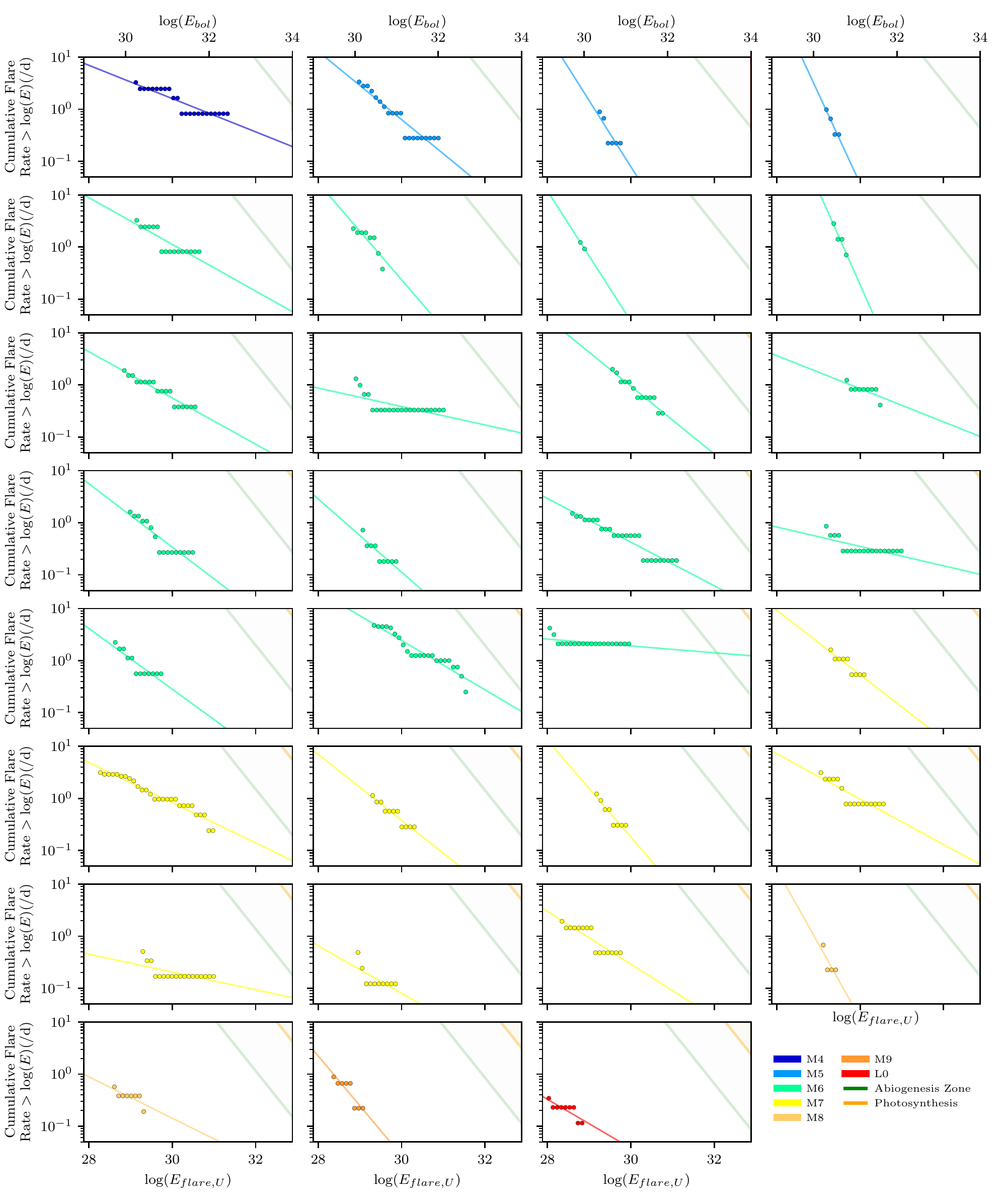}
    \caption{The FFDs for each of the 31 objects with at least three flares. The cumulative flare rates against bolometric flare energies (binned every 10$^{0.1}$\,ergs in log space) are shown by the points as well as the best-fit power law for each star. The plots are ordered in increasing spectral type and the colour of the points and lines correspond to that object's spectral type. The thresholds to reach the abiogenesis zone and perform oxygenic photosynthesis are shown for each object by the green and orange lines respectively.}
    \label{fig:ffd_all}
\end{figure*}

\bsp	
\label{lastpage}
\end{document}